\documentclass[aps,amsmath,prl,twocolumn,superscriptaddress]{revtex4-1}
\usepackage{amsmath,amssymb,graphicx,bm,color,booktabs}
\usepackage{xr}
\usepackage{subfigure}
\usepackage{graphicx}
\usepackage{rotating} 
\usepackage{color}
\usepackage{rotating}
\usepackage{epstopdf}
\usepackage{placeins}
\usepackage{bm}

\vspace*{0.2in}

\begin{document}
\title[Heat dissipation from enzymes]
  {Quantifying the Heat Dissipation from a Molecular Motor's Transport Properties in Nonequilibrium Steady States}

\author{Wonseok Hwang}
\affiliation{Korea Institute for Advanced Study, Seoul 02455, Republic of Korea}
\author{Changbong Hyeon}
\email{hyeoncb@kias.re.kr}
\affiliation{Korea Institute for Advanced Study, Seoul 02455, Republic of Korea}


\begin{abstract}
Theoretical analysis, which maps single molecule time trajectories of a molecular motor onto unicyclic Markov processes, allows us to evaluate the heat dissipated from the motor and to elucidate its dependence on the mean velocity and diffusivity.
Unlike passive Brownian particles in equilibrium, the velocity and diffusion constant of molecular motors are closely inter-related to each other. 
In particular, our study makes it clear that the increase of diffusivity with the heat production is a natural outcome of active particles, which is reminiscent of the recent experimental premise that the diffusion of an exothermic enzyme is enhanced by the heat released from its own catalytic turnover. 
Compared with freely diffusing exothermic enzymes, kinesin-1 whose dynamics is confined on one-dimensional tracks is highly efficient in transforming  conformational fluctuations into a locally directed motion, thus displaying a significantly higher enhancement in diffusivity with its turnover rate. 
Putting molecular motors and freely diffusing enzymes on an equal footing, our study offers thermodynamic basis to understand the heat enhanced self-diffusion of exothermic enzymes. 
\end{abstract}
\maketitle


Together with recent studies \cite{Muddana2010_JACScomm,Yu2009_JACScommu,Sengupta2013JACS,Sengupta2014_Nano}, 
Riedel \emph{et al.} \cite{Riedel2015_Nature} have demonstrated a rather surprising result, from a perspective of equilibrium statistical mechanics: 
Diffusion constants ($D$) of exothermic enzymes, measured from fluorescence correlation spectroscopy (FCS), increase linearly with their catalytic turnover rates $V_{cat}$, so that the enhancement of diffusivity at maximal activity (maximum $V_{cat}$) is as large as $\Delta D/D_0[\equiv (D_{max}-D_0)/D_0]\approx 0.3-3$, where $D_0$ and $D_{max}$ are the diffusion constants measured by FCS at $V_{cat}=0$ and maximal $V_{cat}$, respectively.  
Their enigmatic observation \cite{Riedel2015_Nature} has called much attention of biophysics community to the physical origin of the activity-dependent diffusivity of a single enzyme.   
Golestanian \cite{golestanian2015PRL} considered four distinct scenarios (self-thermophoresis, boost in kinetic energy, stochastic swimming, collective heating) to account for this observation quantitatively; however, the extent of enhancement observed in the  experiment was still orders of magnitude greater than the theoretical estimates from the suggested mechanisms.
Bai \emph{et al.} \cite{bai2015JCP} also drew a similar conclusion by considering hydrodynamic coupling between the conformational change of enzyme and surrounding media. 
The experimental demonstration of enzymes' enhanced diffusion with multiple control experiments in Ref.\cite{Riedel2015_Nature} is straightforward; however, physical insight
of the observed phenomena is currently missing  \cite{Riedel2015_Nature,golestanian2015PRL,Tsekouras_arxiv_2016,Golestanian_arxiv_2016}. 
    
 While seemingly entirely different from freely diffusing enzymes, kinesin-1 \cite{Visscher99Nature,Nishiyama01NCB,Yildiz04Science,Cross05Nature,Yildiz08Cell,Dey2015_Nano} is also a substrate-catalyzing enzyme. 
    Conformational dynamics of kinesin-1, induced by ATP hydrolysis and thermal fluctuations, is rectified into a unidirectional movement with a high fidelity \cite{Hyeon07PNAS2,Zhang2012Structure}; hydrolysis of a single ATP almost always leads to 8 nm step \cite{Schnitzer97Nature}.  
 Every step of kinesin-1 along 1D tracks, an outcome of cyclic chemical reaction, can be mapped onto the chemical state space. Once this mapping is established it is straightforward to calculate the motor velocity ($V$), diffusion constant ($D$), and heat dissipation ($\dot{Q}$) in terms of a set of transition rate constants, thus offering an opportunity to scrutinize the catalysis enhanced-diffusivity of enzymes from a refreshing angle. 
    
    \begin{figure}[t]
    	\centering
    	\includegraphics[scale=0.45]{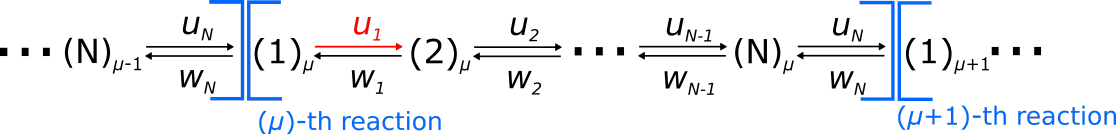}
    	\caption{N-state kinetic model. $(1)_\mu$, $(2)_\mu$, ..., $(N)_\mu$ denote distinct chemical states of a molecular motor going through the $\mu$-th reaction cycle. The microscopic rate constants in the forward ($n\rightarrow n+1$) and backward ($n+1\rightarrow n$) directions are given by $u_n$ and $w_n$, respectively.}
    	\label{fig_cyclic_model}
    \end{figure}

    Here, we map the problem of kinesin-1 onto  Derrida's periodic 1D hopping model (Fig.\ref{fig_cyclic_model}) \cite{Derrida1983_JSP,Fisher99PNAS,Fisher01PNAS} and study the relationship between $D$ and $V$ of the motor. 
    Our study shows that when $V$ is augmented by increasing the substrate (e.g. ATP) concentration, 
    $D$ can be expressed as a third degree polynomial in $V$.  
    Similar to Riedel \emph{et al.}'s measurements on enzymes, the data of kinesins clearly demonstrate the enhanced diffusivity at higher activity (velocity), but the extent of enhancement is even greater.   
    We compare our analyzed result of motor enzyme, kinesin-1, with Riedel \emph{et al.}'s freely diffusing active enzymes and discuss their common feature and differences. From a perspective of thermodynamics, we argue that these two systems belong to the same thermodynamic class in that the dynamics of both systems are affected by the supply of chemical energy input from substrates.
    Our study also clarifies the relationship between the heat dissipation ($\dot{Q}$) and  enhanced diffusivity of the  motor using the theoretical framework of nonequilibrium steady state (NESS) thermodynamics  \cite{oono1998PTPS,Hatano01PRL,Hyeon09PCCP,Qian2005_BiophyChem,Qian07ARPC,Qian2005_BiophyChem,Qian:2000,Qian07ARPC,Qian_PRE_2004,Toyabe:2010bm,Qian:2008:EPL}, and confers thermodynamic insight into how chemical free energy deposited into a molecular system determines its transport properties. 
\\

{\bf Dependence of diffusivity on motor velocity. }
	Kinesin-1 walks along microtubules, hydrolyzing one ATP per step \cite{Schnitzer97Nature,Yildiz04Science,Hyeon11BJ}.
	To model the kinesin's stochastic movement, one can consider a kinetic cycle consisting of $N$ chemically distinct states, where  the probability of being in the $n$-th chemical state, $p_n(t)$ ($n=1,2,\ldots,N$), obeys a master equation (see Fig.\ref{fig_cyclic_model}) \cite{Derrida1983_JSP,Fisher99PNAS,Fisher01PNAS}, 
	\begin{align} \label{eq:masterequation}
	\frac{d p_n(t)}{d t}=u_{n-1}p_{n-1}(t)+w_{n}p_{n+1}(t)-(u_n+w_{n-1})p_n(t)
	\end{align} 
	with $u_{N+n}=u_{n}$, $u_N=u_0$, $w_{N+n}=w_{n}$, $w_N=w_0$, $p_{N+n}(t)=p_{n}(t)$, and $\sum_{n=1}^Np_n(t)=1$.
	Here $p_{n}(t) = \sum_{\mu=-\infty}^{\infty} P_{\mu, n}(t)$ where $P_{\mu, n}(t)$ is the probability of being in the $n$-th chemical state at the $\mu$-th reaction cycle.
	The forward and backward hopping rates between the $n$-th and $(n+1)$-th state are denoted as $u_n$ and $w_n$, respectively.  	
	With $p_n(\infty)=p_n^{ss}$, the steady state flux $j$ along the cycle is expressed as follows, 
	\begin{align}
	j=u_np_n^{ss}-w_np_{n+1}^{ss}=\frac{\prod_{n=1}^Nu_n-\prod_{n=1}^Nw_n}{\Sigma(\{u_n\},\{w_n\})}
	\end{align}
	where $\Sigma(\{u_n\},\{w_n\})\equiv \prod_{n=1}^Nu_n\sum_{n=1}^Nr_n$ with 
	$r_n=u_n^{-1}\left[1+\sum_{i=1}^{N-1}\prod_{j=1}^i(w_{n+j-1}/u_{n+j})\right]$ \cite{Derrida1983_JSP}.
	The net flux $j$ is decomposed into the forward and backward fluxes as, $j=j_+-j_-$ where $j_+=\prod_{n=1}^Nu_n/\Sigma(\{u_n\},\{w_n\})$ and $j_-=\prod_{n=1}^Nw_n/\Sigma(\{u_n\},\{w_n\})$.
	Both the mean velocity ($V$) and effective diffusion constant ($D$) are defined from the \emph{trajectories}, $x(t)$, that record the position of individual kinesin motors:  
	  
\begin{align}
V=\lim_{t\rightarrow\infty}\frac{d}{dt}\langle x(t)\rangle=d_0\lim_{t\rightarrow\infty}\frac{d}{dt}\left[\sum_{\mu=-\infty}^{\infty}\mu \pi_\mu(t)\right]
\end{align} 
and 
\begin{align}
D&=\frac{1}{2}\lim_{t\rightarrow\infty}\frac{d}{dt}\left(\langle (x(t))^2\rangle-\langle x(t)\rangle^2\right)\nonumber\\
&=\frac{d_0^2}{2}\lim_{t\rightarrow\infty}\frac{d}{dt}\left[\sum_{\mu=-\infty}^{\infty}\mu^2 \pi_\mu(t)-\left(\sum_{\mu=-\infty}^{\infty}\mu \pi_\mu(t)\right)^2\right],
\label{eqn:D}
\end{align}
where $\pi_{\mu}(t) = \sum_{n=1}^{N} P_{\mu,n}(t)$, and $d_0$ is the step size.
Both $V$ and $D$ are fully determined in terms of a set of rate constants, $\{u_n\}_{n=1,\ldots N}$ and $\{w_n\}_{n=1,\ldots, N}$ \cite{Fisher99PNAS,Fisher01PNAS} (See SI). 
Regardless of the nature of dynamical process (equilibrium or non-equilibrium, passive or active, biased or unbiased), the first line of Eq.\ref{eqn:D} is the general definition of diffusion constant. 
Most experiments directly calculate the value of $D$ from trajectories based on Eq.\ref{eqn:D}, or at least extract the value of $D$ from formulae derived based on Eq.\ref{eqn:D} (e.g. auto-correlation function of FCS by assuming the normal diffusion \cite{Riedel2015_Nature}).

	When [ATP] is the only control variable, a simple relationship between $V$ and $D$ is derived by assuming that $u_1(=u_1^o[\text{ATP}])$ is the only ATP-dependent step in the reaction scheme (Fig.\ref{fig_cyclic_model}). 
		Since $V$ and $D$ are both functions of $[\text{ATP}]$ \cite{Fisher99PNAS,Fisher01PNAS}, 
	it is possible to eliminate the common variable $[\text{ATP}]$ (or more conveniently $u_1$) from the two quantities. 
	For the general $N$-state model, one can express $D$ as a third degree polynomial in $V$ (see  SI for $N=1$, 2, and the details of derivation for the $N$-state model):
		\begin{equation}
			D(V) = D_0+  \alpha_1 V- \alpha_2 V^2 + \alpha_3 V^3. 
		\label{eq:D_V}
		\end{equation}
		where $\alpha_i$'s are the constants, uniquely defined when all the rate constants $\{u_n\}_{n=2,\ldots N}$ and $\{w_n\}_{n=1,\ldots, N}$ are known.
		
		This relationship (Eq.\ref{eq:D_V}) holds as long as a motor particle retaining $N$ internal chemical states walks along 1D tracks which are made of binding sites with an equal spacing. 
		In fact, the enhancement of diffusion in motor particles has also been noted by Klumpp and Lipowsky \cite{Klumpp:PRL:2005} in the name of active diffusion and a similar form of velocity dependent diffusion constant as Eq.\ref{eq:D_V} was obtained. 
		The detail of their expression differs from Eq.\ref{eq:D_V}, however, because the focus of their study was on the effect of the patterns (or geometry) of underlying scaffold on the active diffusion constant of the motor. 

	Eq.\ref{eq:D_V} was used to fit the ($V$,$D$) data digitized from Visscher \emph{et al.}'s single molecule measurement on kinesin-1 \cite{Visscher1999_Nature} which had reported $V$ and the randomness parameter $r = 2 D/d_0 V$ ($d_0 = 8.2$ nm, kinesin's step size) at varying load ($f$) and [ATP].     
\begin{figure}[ht]
	\centering
	\includegraphics[scale=0.4]{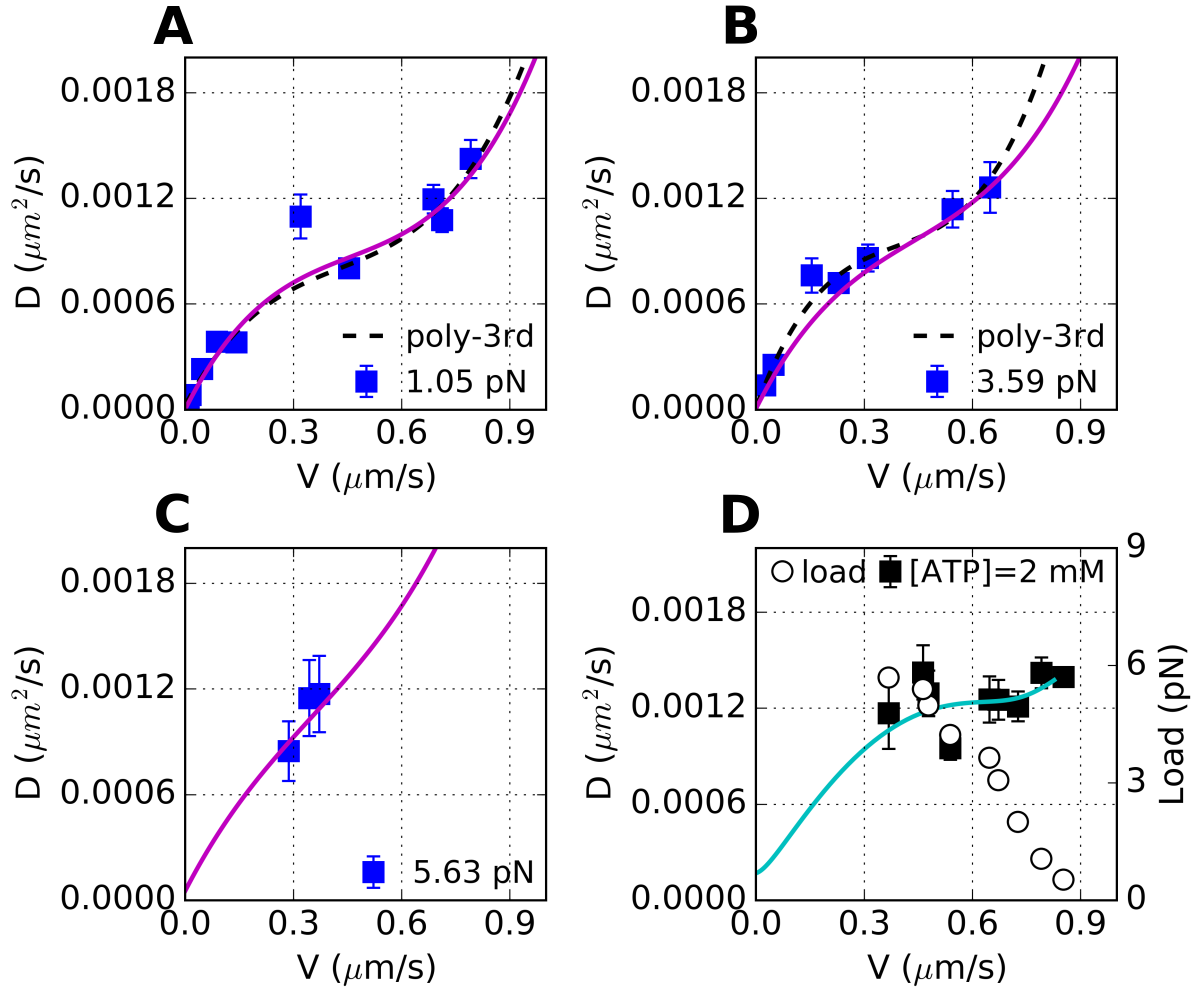}
	\caption{Motor diffusivity ($D$) as a function of mean velocity ($V$) of kinesin-1.  
		($V$,$D$) measured at varying [ATP] (= 0 -- 2 mM) and a fixed ({\bf A}) $f=$1.05 pN, ({\bf B}) 3.59 pN, and ({\bf C}) 5.63 pN \cite{Visscher1999_Nature}. The standard deviations of $D$ ($\sigma_{D}$) were estimated from $\sigma_{D} \simeq d_0 (\sigma_{r} V + r\sigma_{V})$ by using the extracted values of $r$, $V$, $\sigma_{r}$ and $\sigma_V$.
		The black dashed lines in {\bf A} and {\bf B} are the fits using Eq.\ref{eq:D_V}.
		For $f=1.05$ pN and 3.59 pN, $(D_0,\alpha_1,\alpha_2,\alpha_3)=(2.2\times 10^{-5},3.8\times 10^{-3},7.1\times 10^{-3},5.5\times 10^{-3})$, and $(7.4\times 10^{-6},5.6\times10^{-3},1.2\times 10^{-2},1.1\times 10^{-2})$, respectively.
		The solid lines in magenta in {\bf A-C} are plotted using the ($N$=4)-kinetic model's parameters (Table \ref{table}).
		{\bf D.} ($V$,$D$) (black filled square) measured at varying $f$ (black empty circle) and [ATP] = 2 mM.
		The solid line in cyan, plotted by using the parameters in Table \ref{table}, is the predicted behavior of $D=D(V)$ when $V$ is varied by $f$, instead of [ATP]. 
	}
	\label{fig_VD}
	
\end{figure}  
The fits (dashed line) using Eq.\ref{eq:D_V} allow us to determine the parameters, $D_0$, $\alpha_1$, $\alpha_2$, and $\alpha_3$ (see Fig.\ref{fig_VD}A ($f=1.05$ pN) and Fig.\ref{fig_VD}B ($f=3.59$ pN)).    
As expected, $D(V=0)=D_0\approx 10^{-5}$ $\mu$m$^2$/sec is vanishingly small for kinesin-1 whose motility is tightly coupled to ATP. 
At $V=0$, the flux along the cycle vanishes ($j=0$), establishing the detailed balance (DB), $u_np_n^{eq}=w_np_{n+1}^{eq}$ for all $n$'s with $\sum_{n=1}^Np_n^{eq}=1$. 
 In this case, $D_0=d_0^2/\sum_{n=1}^N(u_np_n^{eq})^{-1}\leq u_{\text{min}}d_0^2/N$ \cite{kalnin2013JCP}, where $u_{\text{min}}=\min{\{u_n|n=1\ldots N\}}$.   
For $[\text{ATP}]\ll 1$, it is expected that $u_{\text{min}}\approx u_1=u_1^o[\text{ATP}]\ll 1$. 
 
\begin{figure}[ht]
	\centering
	\includegraphics[scale=0.34]{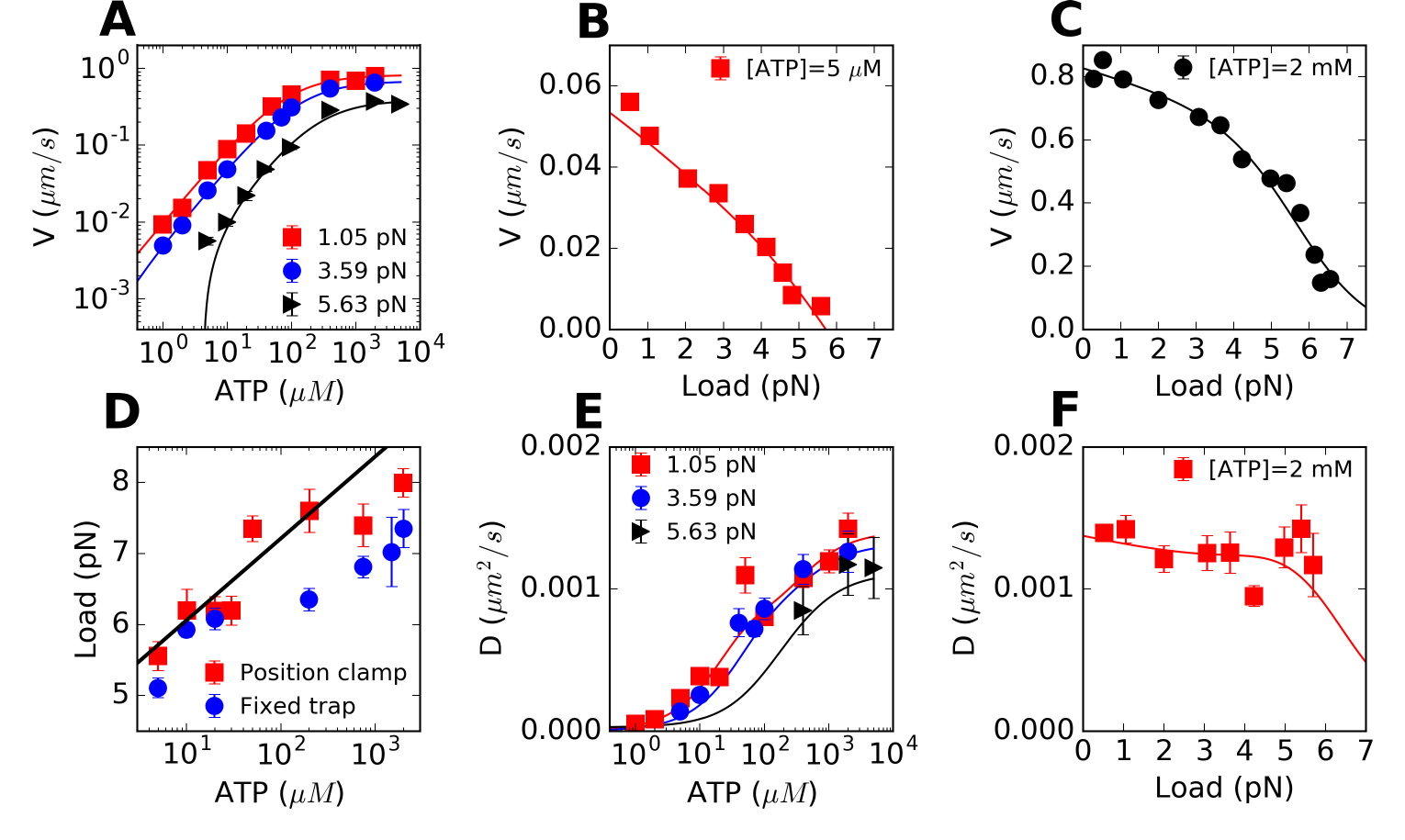}
	
	\caption{
		Analysis of experimental data, digitized from Ref. \cite{Visscher1999_Nature}, using ($N$=4)-state cyclic model.
		The solid lines are the fits to the data
		{\bf A.} $V$ vs [ATP] at $f=$1.05 pN (red square), 3.59 pN (blue circle), and 5.63 pN (black triangle).
		{\bf B.} $V$ vs $f$ at [ATP] = 5 $\mu$M.
		{\bf C.} $V$ vs $f$ at [ATP] = 2 mM.
		{\bf D.} Stall force as a function of [ATP], measured by `Position clamp' (red square) or `Fixed trap' (blue circle) methods.
		{\bf E.} $D$ vs [ATP] at $f=$1.05 pN (red square), 3.59 pN (blue circle), and 5.63 pN (black triangle).
		$D$ was estimated from $r=2D/Vd_0$. 
		{\bf F.} $D$ vs $f$ at [ATP] = 2 mM.
	}
	\label{fig_kinesin_refit_fix}
\end{figure}

	We also used the ($N$=4)-state kinetic model by 
	Fisher and Kolomeisky \cite{Fisher01PNAS} and determined a set of parameters, $\{u_n\}$, $\{w_n\}$, and $\{\theta^{\pm}_n\}$ (with $n=1,\ldots 4$),   
	which best describe the kinesin's motility data, by simultaneously fitting all the data points in Fig.\ref{fig_VD}A-C and Fig.\ref{fig_kinesin_refit_fix} (see ``Analysis of kinesin-1 data using (N=4)-state kinetic model'').
	For a consistency check, we overlaid a theoretically predicted line (Fig.\ref{fig_VD}D, cyan line) over the data ($V$,$D$) obtained at varying $f$ but with fixed [ATP]=2 mM, which we did not use in determining the parameters. 
	$D(V)$, over the range of $0<V \lesssim 0.3$ $\mu$m/sec (Fig.\ref{fig_VD}D), predicts the behavior of $D$ at high $f$ regime near a stall force.
\\

\begin{figure*}[t]
	\centering
	\includegraphics[scale=0.45]{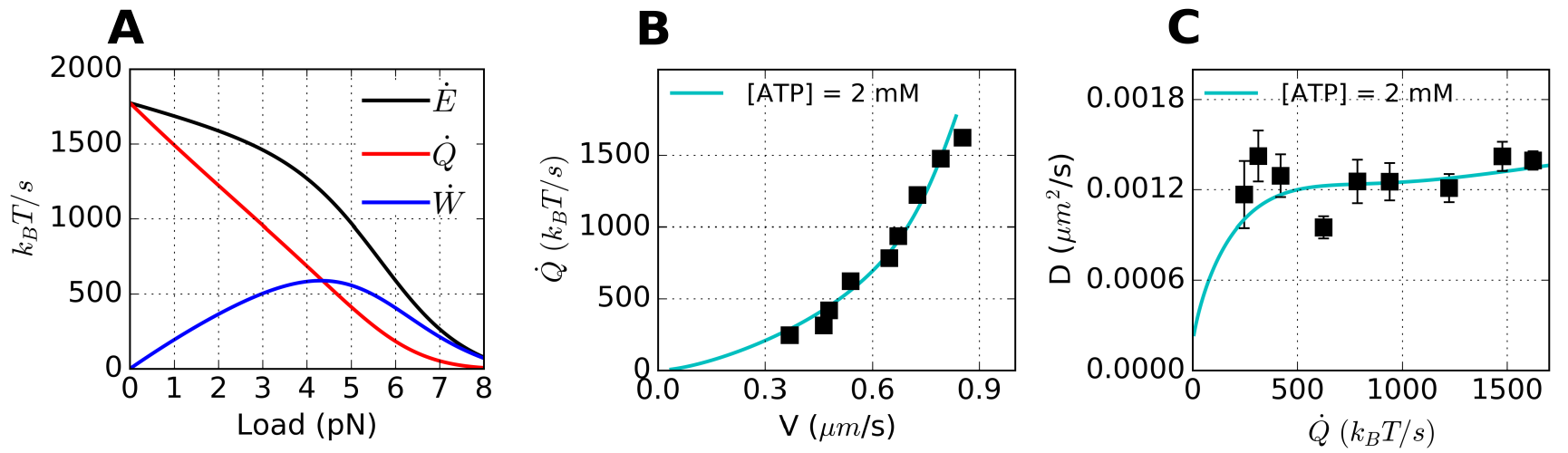}
	\caption{
		Heat and work production at varying load.
		{\bf A.} Theoretical plot of heat ($\dot{Q}$, red) and work production ($\dot{W}$, blue), and their sum ($\dot{E}$, black) as a function of load using ($N$=4)-state model.
		{\bf B.} Heat production ($\dot{Q}$) as a function of motor velocity ($V$), modulated by varying $f$ at [ATP] = 2 mM.
		{\bf C.} $D$ plotted against $\dot{Q}$ when $f$ is varied at [ATP] = 2 mM.
		The solid lines in cyan are theoretical predictions using the parameters determined in the ($N$=4)-state model.
	}
	\label{fig_Q_load}
\end{figure*}  
{\bf Energy and heat balance of molecular motor. }
	The movement of a molecular motor is driven by a net driving force due to ATP hydrolysis and opposed by the resisting load $f$.  
	In a NESS, 
	the flux ratio, $K(f)=j_+(f)/j_-(f)$, defined for unicyclic reaction cycle for kinesin,  
	is balanced with the chemical potential difference driving the reaction $\Delta\mu_{\text{eff}}(f)$ (or the affinity $\mathcal{A}=-\Delta\mu_{\text{eff}}$) as 
	\begin{align}
	K(f)=\frac{\prod_{n=1}^Nu_n(f)}{\prod_{n=1}^Nw_n(f)}=\exp{\left(-\Delta\mu_{\text{eff}}(f)/k_BT\right)},
	\end{align}
	where 
	$\Delta \mu_{\text{eff}}$ 
is contributed by chemical potential due to ATP hydrolysis $\Delta \mu_{\text{hyd}}$ and mechanical work ($fd_0$) against the load $f$. 
	With $j(f)$ denoting the total flux (i.e., the number of cycles per a given time) at force $f$,  
	the heat dissipated at a steady state, $\dot{Q}=j(f)\times (-\Delta\mu_{\text{eff}})$, is 
	balanced with the (free) energy consumption $\dot{E}=j(f) \times (- \Delta\mu_{\text{hyd}})$ subtracted by the work against an external load $\dot{W}=j(f)fd_0$, and thus   
	\begin{align}
	\dot{Q}&=j(f)\times (-\Delta\mu_{\text{eff}}(f))\nonumber\\
	&=(j_+(f)-j_-(f))k_BT\log{\left(\frac{j_+(f)}{j_-(f)}\right)}\nonumber\\
	&=j(f)\times ( - \Delta\mu_{\text{hyd}} -fd_0)=\dot{E}-\dot{W}
\label{eq:Qtot}
	\end{align} 
	where $\dot{Q}$, analogous to the electric power produced by means of current$\times$voltage, is always positive ($\dot{Q}\geq 0$) regardless of whether $j(f)>0$ or $j(f)<0$.  
        Eq.\ref{eq:Qtot} is readily obtained by assuming barometric dependence of rates on forces as $u_n = u_n^o e^{-f d_0 \theta_n^+ / k_B T}$ and $w_n = w_n^o e^{f d_0 \theta_n^- / k_B T}$ with $\sum_{n=1}^N\left( \theta_n^+ + \theta_n^- \right) = 1$ \cite{Fisher99PNAS,Fisher01PNAS}. 
        When $f=0$, the motor moves along microtubules uni-directionally 
        but the movement of motor itself does not perform work to the environment; 
        thus, the entire free energy consumed via ATP hydrolysis (-$\Delta\mu_{\text{hyd}}>0$) is dissipated into  heat at a rate $j(0)\times(-\Delta\mu_{\text{hyd}})$. 
        When $f\neq 0$, the motor performs work against the load, $W= f d_0$ per cycle. 
        Hence, the total chemical free energy  change due to ATP hydrolysis $-\Delta \mu_{\text{hyd}}$ 
         is dispensed into heat ($Q$) and work ($W$) per cycle, leading to $\dot{E} = \dot{Q} + \dot{W}$ \cite{Toyabe:2010bm}.
         Note that $\dot{W}=j(f)fd_0=0$ either at $f=0$ or at the stall condition $f=f_c$ which imposes $j(f_c)=0$; 
	thus the work production ($\dot{W}$) is a non-monotonic function of $f$, whereas $\dot{E}$ and $\dot{Q}$ decrease monotonically with $f$.
	For concreteness, we plot $\dot{E}$, $\dot{Q}$ and $\dot{W}$ as a function of $f$ (Fig.\ref{fig_Q_load}A). 
	At [ATP] = 2 mM, $\dot{W}$ is maximized at $f \approx 4.5$ pN.
	The heat production $\dot{Q}$ is maximal $\approx 1750$ $k_BT/s$ at $f=0$, and decreases monotonically to zero at stall ($f=f_c$).

	The monotonic increase of $\dot{Q}(V)$ (Fig.\ref{fig_Q_load}B) implies that more heat is generated when the motor moves faster at a smaller $f$. 
	Higher load ($f$) that hampers motor movement (smaller $V$) as in Fig.\ref{fig_VD}D reduces $\dot{Q}$ (Fig.\ref{fig_Q_load}B).
	If the dissipated heat does influence the dispersion of motor, then a positive correlation between $\dot{Q}$ and $D$ should be observed even when both quantities are suppressed at higher force. 
	Indeed, Fig.\ref{fig_Q_load}C predicts that $D$ increases with $\dot{Q}$, although the extent of the increase is small over the range where the data are available.

\begin{figure}[b]
	\centering
	\includegraphics[scale=0.4]{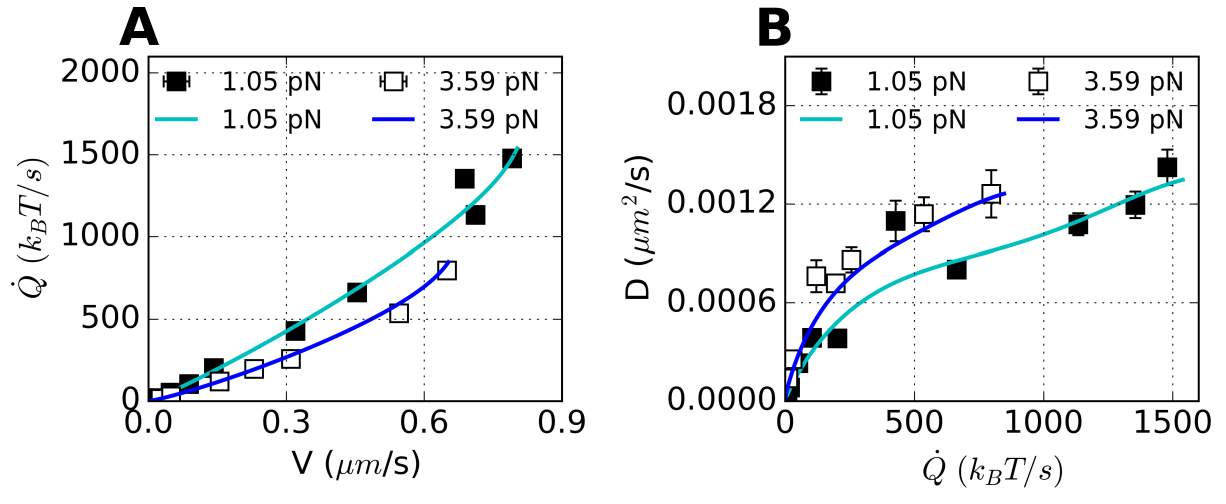}
	\caption{
		The relationships of $\dot{Q}$ vs $V$, and $D$ vs $\dot{Q}$ modulated under varying [ATP] but at a constant $f$. 
		{\bf A.} $\dot{Q}$ vs $V$ at $f=1.05$, 3.59 pN. 
		{\bf B.} $D$ vs $\dot{Q}$ at $f=1.05$, 3.59 pN.
		Solid lines are the fits using ($N$=4)-state model with model parameters determined from global fitting of data in Fig.\ref{fig_VD}A-C, \ref{fig_kinesin_refit_fix}.
		The $V$ and $D$ data are digitized from \cite{Visscher1999_Nature}.
	}
	\label{fig_Q_ATP}
\end{figure}

	Next, to investigate the effect of varying [ATP] on $V$, $D$, and $\dot{Q}$,   
	we plotted ($V$,$\dot{Q}$) (Fig. \ref{fig_Q_ATP}A) and ($\dot{Q}$,$D$) (Fig.\ref{fig_Q_ATP}B) at varying [ATP] with fixed $f=1.05$ pN or $f=3.59$ pN. 
	Again, monotonic increase of $\dot{Q}$ with $V$, and  the correlation between $\dot{Q}$ and $D$ clearer than that in Fig.\ref{fig_Q_load}C are observed.
	Similar to the cubic polynomial dependence of $D$ on $V$, 
	it is possible to relate $V$ and $D$ with $\dot{Q}$ at constant load. 
	We found that for general $N$-state model $\dot{Q}\sim V^2$ and $D\sim \dot{Q}^{1/2}$ at small $\dot{Q}$ (see SI section 5), which explain the curvatures of the plots at small $\dot{Q}$ regime in Fig.\ref{fig_Q_ATP}. 
From the perspective of NESS thermodynamics \cite{oono1998PTPS,Hatano01PRL}, for a motor to sustain its motility, a free energy cost called \emph{housekeeping heat} should be continuously supplied to the system. 
For the $N$-state model, the system relaxes to the NESS from its arbitrary initial non-equilibrium state in a rather short time scale $\tau_{\text{NE}} =1/\sum_{n=1}^N(u_n+w_n)$ (see SI section 4 and Fig.S2). 
In a NESS, the housekeeping heat, and the total heat and entropy production discharged to the heat bath are all equal to $\dot{Q}=jk_BT\log{(j_+/j_-)}\geq 0$ (see SI section 4).

Cautionary remarks are in place. 
Our formalism describing the trajectories of kinesin is solely based on a unicyclic reaction scheme. 
While straightforward in developing a formalism, the unicyclic reaction scheme leads to a problematic interpretation that the backstep is realized always by a reversal of the forward cycle \cite{Hyeon09PCCP}, which means that the backstep near stall condition is taken with the synthesis of ATP from ADP and P$_i$.   
	This rather strong assumption could be alleviated by extending the current formalism to the one based on a multi-cycle model \cite{Liepelt07PRL,Yildiz08Cell,Hyeon09PCCP,clancy2011NSMB}, so as to accommodate the possibilities of ATP-induced backstep and futile cycle near the stall condition. 
	For multiple cycle model, the flux branches into different cycles and the net flux at each kinetic step remains nonvanishing ($j_+\neq j_-$) even at stall condition. 
	As a result, it is expected that $\dot{Q}\neq 0$ and $\dot{W}\neq 0$. 
	More explicit calculation of the functional dependence of $\dot{Q}$ or $\dot{W}$ on $f$, however, requires a detailed model based on multi-cycle reaction scheme, which we leave for our future study.  
	\\

{\bf Passive versus active particles.} Broken DB and violation of FDT \cite{Battle2016Science,Gladrow2016PRL} differentiate an active system operated under non-conservative forces from a passive system in mechanical equilibrium under conservative forces. 
    For example, the terminal velocity ($V$) and diffusion constant ($D$) of a colloidal particle of size $R$ in 
    the gravitational or electric field, are mutually independent, so that regardless of $V$, $D$ is always constant, obeying Stokes-Einstein relation $D\sim D_{\text{SE}}\sim k_BT/\eta R$ where $\eta$ is the viscosity of media, $k_B$ is the Boltzmann constant, and $T$ is the absolute temperature. 
    A similar argument can be extended to a composite system (e.g., macromolecules in solution) subjected to conservative forces.
    
    In contrast, for a self-propelled active particle,  
    the dependence of diffusivity on its velocity is often noted, and the \emph{effective diffusion constant}, defined as the increment of mean square displacement over time $D_{\text{eff}}=\langle(\delta r)^2\rangle/6t$ at an ambient temperature $T$, depends on a set of parameters (velocity, density, etc.), violating the FDT \cite{Tailleur_2008_PRL,Liu2011Science}. 
    To be specific, let us consider a run-and-tumble motion of a swimming bacterium, which locomotes with a velocity $V_b$ in search of a food. 
    If the mean duration of locomotion is $\tau_r$ and the bacterium tumbles occasionally with a rotational diffusion constant $D_R$ for time $\tau_t$, 
    the effective diffusion constant of the bacterium at time $t$ much greater than $\tau_s$ and $\tau_t$ is estimated $D_{\text{eff}}\sim V_b^2\tau_r/6D_R\tau_t$ \cite{Condat_2005_PRE,Lovely_1975_JTheoBiol}.
    In this case, $V_b$ or $D_{\text{eff}}$ of bacterium is affected not by the ambient temperature but by the amount of food, also violating the conventional FDT ($D_{\text{eff}}\nsim k_BT/\eta R$) \cite{mizuno2007Science,marchetti2013RMP,Battle2016Science}.
    
    Unlike a passive particle in equilibrium, $V$ and $D$ of an active particle are both augmented by the same \emph{non-thermal}, \emph{non-conservative} force (e.g. ATP hydrolysis).  
Importantly, regardless of whether a system is in equilibrium or in non-equilibrium, 
and is passive or active, it is legitimate to \emph{define} the diffusion constant as an increased amount of mean square displacement for time $t$ \emph{without} resort to the fluctuation dissipation theorem (FDT). 
In Ref. \cite{Riedel2015_Nature} the signal from FCS measurement was nicely fitted to the auto-correlation function $G(\tau)$ which assumed  the \emph{normal} diffusive motion of the enzymes. 
\\

{\bf Comparison of enhanced diffusivities between different types of active particles. }
While a precise \emph{mechanistic link} between the heat and enhanced diffusion is still elusive in this study as well as in others 
\cite{bai2015JCP,Riedel2015_Nature,golestanian2015PRL,Tsekouras_arxiv_2016,Golestanian_arxiv_2016},  
our study still offers further insights into the problem of enhanced diffusion of exothermic enzymes \cite{Muddana2010_JACScomm,Riedel2015_Nature}. 
From Fig.\ref{fig_VD}, 
$(\Delta D/D_0)_{\text{obs}}$ at the maximal velocity of kinesin-1 is as large as $\sim \mathcal{O}(10^2)$. 
For swimming \emph{E. coli} the enhancement is estimated $(\Delta D/D_0)_{\text{obs}}\gtrsim  \mathcal{O}(10^2)$ (the effective diffusion coefficient of \emph{E. coli} is $D \sim 53$ $\mu m^2 /s$ \cite{Wu_2006,Condat_2005_PRE} and $D_0=D_{\text{SE}} \sim 0.5$ $\mu m^2/s$ by assuming bacterium as a sphere with radius of $0.5$ $\mu m$). 
Considering the extents of enhancement in kinesin-1 and \emph{E. coli}, 
$(\Delta D/D_0)_{\text{obs}}\sim 0.3-3$ for the substrate fed, exothermic enzymes observed by Riedel \emph{et al.} \cite{Riedel2015_Nature} should not be too surprising.

\begin{table*}[th!]
	\caption{Rate constants, enhancement of diffusion, and conversion factor determined from the (N=2)-state kinetic model. 
		AP: Alkaline phosphatase, TPI: Triose phosphate isomerase. 
		$^{\dagger}$ $D_0$ determined from the 3rd degree polynomial fit (Eq.\ref{eq:D_V}) to the data in Fig.\ref{fig_VD}A was used to estimate $(\Delta D/D_0)_{\text{obs}}$ and $\psi$. 
	}
	\scalebox{0.98}{
		\begin{tabular}{ | c ||    c   |    c   |   c    |    c    |   c    |    c   ||  c  |  c ||  c  |}
			\hline
			& $Q$ ($k_B T$) &  [S] (mM) & $u_1$ ($s^{-1}$) & $u_2$ ($s^{-1}$) & $w_1$ ($s^{-1}$)  & $w_2$ ($s^{-1}$) &  $ \left( \frac{\Delta D}{ D_0}\right)_{\text{obs}} $  & $\left(\frac{\Delta D}{D_0}\right)_{\text{max}}$ & $\psi$ \\ \hline\hline
			kinesin ($f=1.05$ pN)	    &  15 &  2 & 2200  &  99 & 0.55 & 0.092 &  $6\times 10^4$ ($^{\dagger}45$) & $9.7 \times 10^4$ & $\sim 0.8$ $(^{\dagger}0.02)$ \\ 
			\hline\hline
			Catalase   & 40  & 62 & $6.2 \times  10^6$  &  $5.8\times10^4$ &  $6.1 \times 10^6$ & $2.2 \times 10^{-13}$&  $\sim 1 $  & $ 1.3 \times 10^{17}$ & $\sim 3 \times 10^{-9} $ \\\hline
			Urease   & 24  & 3 & $3 \times  10^5$  &  $1.7\times10^4$ &  $2.8 \times 10^5$ & $7.4 \times 10^{-7}$ &  $\sim 0.3 $  & $ 1.2 \times 10^{10}$ & $\sim 5 \times 10^{-6} $\\\hline
			AP   & 17  & 1.6 & $1.6 \times  10^5$  &  $1.4\times10^4$ &  $1.5 \times 10^5$ & $4.0 \times 10^{-4}$&  $\sim 3 $  & $ 1.9 \times 10^{7}$ & $\sim  4\times 10^{-4} $\\\hline
			TPI  & 1.2  & 1.8 & $1.8 \times  10^5$  &  $1.3\times10^4$ &  $1.7 \times 10^5$ & $4.2 \times 10^{3}$&  0.01 & 1.2 & $ 0.09 $\\
			\hline
		\end{tabular} 
	}
	\label{table2}
\end{table*}

In the framework of unicyclic Markov processes, the diffusion constant ($D$) in a NESS is defined consistently with Eq.\ref{eqn:D} in terms of forward and backward fluxes ($j_+$ and $j_-$). 
The extent of enhancement in diffusion constant is expressed as (see Eq.S41) 
\begin{align}
\frac{\Delta D}{D_0}&=\frac{j_+-j_-}{j_0\log{\left(\frac{j_+}{j_-}\right)}}-1=\left(\frac{j_-}{j_0}\right)\frac{(K-1)}{\log{K}}-1. 
  \label{eqn:enhancement}
\end{align} 
At equilibrium, when the DB is established, $j_+=j_-=j_0$ (or $K=1$), which leads Eq.\ref{eqn:enhancement} to $\Delta D/D_0=0$.   
More explicitly, the enhancement of diffusion constant can be expressed in terms of microscopic rate constants using the (N=2)-state kinetic model (see Eq.S9 in SI section 1) and its theoretical upper bound can be obtained as 
\begin{align}
\frac{\Delta D}{D_0}
  \leq \left(\frac{\Delta D}{D_0}\right)_{\text{max}}=\frac{u_2^2+(w_1+w_2)u_2-w_1w_2}{2w_1w_2}. 
  \label{eqn:enhancement_N2}
\end{align} 
The inequality in the last line specifies 
a theoretically achievable upper bound of enhancement $(\Delta D/D_0)_{\text{max}}$, the expression of which remains unchanged even when the passive diffusion component ($D_{\text{SE}}\sim k_BT/\eta R$) is included in $D_0$. 
For a Michaelis-Menten type enzyme reaction, a typical condition, $u_2\gg w_2$ and $u_2\simeq w_1$, makes $(\Delta D/D_0)_{\text{max}}\simeq u_2^2/2w_1w_2$ a large number. 
$D$ (or $D_0$) itself is a number associated with a squared length scale $d_0^2$ per unit time.  
However, the precise meaning of $d_0$, a characteristic length, is not clear for the freely diffusing enzymes while $d_0$ simply denotes the step size for molecular motors.  
The dimensionless number, $(\Delta D/D_0)$, eliminates such ambiguity, allowing us to make a direct comparison between 1D transport motors and enzymes.

In the expression $(\Delta D/D_0)_{\text{max}}\simeq u_2^2/2w_1w_2$, $u_2$ is the key reaction rate that quantifies the catalytic event in Michaelis-Menten scheme (or ``power stroke'' in molecular motors). 
In order to quantify the enzyme's efficiency of converting chemical free energy into motion
we define the conversion factor $\psi$ as the ratio between the \emph{observed} and \emph{theoretically predicted} enhancement of diffusion constant at the maximal turnover rate ($V=V_{max}$) as follows:  
\begin{align}
\psi^2 \simeq\frac{\left(\frac{\Delta D}{D_0}\right)_{\text{obs}}}{\left(\frac{\Delta D}{D_0}\right)_{\text{max}}}. 
\end{align}
Mathematically, the factor $\psi$ amounts to the ratio of $u_2^{\text{obs}}/u_2$ where $u_2^{\text{obs}}$ is an actual amount of power stroke; and hence 
it physically quantifies the extent of chemical energy converted to spatial movement. 
For kinesin-1 whose ATP-induced conformational dynamics and thermal fluctuations are rectified to a unidirectional movement along a 1D track \cite{Schnitzer97Nature}, 
a high conversion factor ($\psi\lesssim 1$), i.e., tight coupling between the transitions in chemical state space and motion in real space is expected from the catalytic turnover. 
In contrast, the lack of scaffold renders the motion of free enzymes in 3D space random and more dissipative, and hence the transitions in chemical state space is weakly coupled to the motion in real space.
As a consequence, the extent of conversion from chemical energy to the movement of enzyme is expected to be much lower than that of kinesin-1. 

Indeed, we find that $\psi$(kinesin) $\gg$ $\psi$(freely diffusing enzymes). 
For kinesin-1, 
$D_{\text{max}}\approx 10^{-3}$ $\mu$m$^2$/s at $V=V_{\text{max}}$ from Fig.\ref{fig_VD} and $D_0 = 10^{-8}$ $\mu$m$^2$/s from the fit to $(N=2)$-state model (see SI) which determines the rate constants $u_2$, $w_1$ and $w_2$ lead to 
$(\Delta D/D_0)_{\text{obs}}\approx 6\times 10^4$ and $(\Delta D/D_0)_{\text{max}}\approx 7.4\times 10^5$ from Eq.\ref{eqn:enhancement_N2}; therefore, $\psi\approx 0.8$ (or $\psi \approx 0.02$ when $D_0\approx 2.2\times 10^{-5}$ $\mu$m$^2$/s is used from the third degree polynomial fit: dashed line in Fig.\ref{fig_VD}A).    
For the cases of Riedel \emph{et al.}'s exothermic enzymes (catalase, urease, alkaline phosphatase), whose rate constants are available in Table \ref{table2} (or in Ref.\cite{Riedel2015_Nature}), $\psi\sim\mathcal{O}(10^{-4})-\mathcal{O}(10^{-7})$ is obtained from 
$(\Delta D/D_0)_{\text{obs}}\sim\mathcal{O}(10^{-1})$ and 
$(\Delta D/D_0)_{\text{max}}\sim\mathcal{O}(10^7)-\mathcal{O}(10^{17})$.

The net chemical free energy change due to isomerization reaction of substrate (dihydroxyacetone phosphate $\rightleftharpoons$ D-glycealdehyde 3-phosphate) catalyzed by triose phosphate isomerase would be relatively small ($\Delta \mu_{\text{eff}}\sim0$ or $K\sim 1$) compared with other highly exothermic enzymes. 
In this case, it is anticipated from the first line of Eq.\ref{eqn:enhancement} that $\Delta D/D_0\sim 0$. 
All the values of $(\Delta D/D_0)_{\text{obs}}$, $(\Delta D/D_0)_{\text{max}}$, and $\psi$ discussed here are provided in Table \ref{table2}.

Direct comparison of the diffusions of kinesin-1 and freely diffusing active enzymes may not appear to be fair. 
From a perspective of thermodynamics, however, they still belong to the same thermodynamic class in that the motions of both systems requires energy input.  
Furthermore, when mapped on the chemical state space, (enzymatic) activities of both systems are described using Michaelis-Menten relation with ATP concentration. 
As quantified in the relation of $\psi$(kinesin-1)$\gg\psi$(freely diffusing enzymes), 
kinesin-1, whose fluctuations are tightly confined on the microtubules, is more efficient in converting thermal/active fluctuations into motion than the freely diffusing enzymes. 
Thus, our prediction is that confinement of active fluctuations into low dimension leads to a greater enhancement in diffusivity $(\Delta D/D_0)_{\text{obs}}$, which can be tested for the above-mentioned freely diffusing enzymes by confining them in a narrow nanochannel. 
Conversely, it is also expected that $(\Delta D/D_0)_{\text{obs}}$ and $\psi$ of free kinesin-1 in solution, i.e., in the \emph{absence} of microtubules, are reduced greatly to the values less than those for Riedel \emph{et al.}'s enzymes. 
\\

{\bf Concluding Remarks.}
The physical meaning of the term ``diffusion constant'' used in the literature could be twofold.  
First, it refers to the response of a system in a solution to thermal fluctuations, which amounts to the diffusion constant defined by the Stokes-Einstein relation, $D_{\text{SE}} = k_B T / \zeta $ where $\zeta$ is the friction coefficient. 
Second, the behavioral random motion of a system being probed is often quantified using the operational definition of diffusion constant 
$D_{\text{eff}}=\langle(\delta r)^2\rangle/6t$ at long time limit. 
In a non-driven thermally equilibrated system, it is expected that $D_{\text{eff}}=D_{\text{SE}}$. 
But, for a system like swimming bacterium, where unidirectional active motion is randomized with occasional tumblings, there is no reason to expect that the two distinct definitions are inter-related, and $D_{\text{eff}}>D_{\text{SE}}$ should be expected as long as the bacterium is ``alive.''
It is important to note that in Riedel \emph{et al.}'s FCS measurement, the behavioral random motion of enzymes was effectively quantified as the diffusivity of the enzymes based on the definition of $D_{\text{eff}}[=\langle(\delta r)^2\rangle/6t]$ 
and its variation with an increasing turnover rate was extracted from the data fitting to the fluorescence intensity auto-correlation function.  
Once one accepts that substrate-catalyzing, freely diffusing enzymes are 
thermodynamically in the same class with molecular motors or swimming bacteria in that all of them are energy-driven (substrate-catalyzing or nutrient-digesting) systems in NESS, 
the enhancement of enzyme diffusion is no longer enigmatic.

The fundamental difference between passive and active particles is worth highlighting again using Langevin description.   
In the simplest possible term, the motion of a passive particle in 1D under an \emph{externally} controlled field, $F_{\text{ext}}$, is described by Langevin equation $\dot{x}(t)=F_{\text{ext}}/\gamma+\sqrt{2D}\zeta(t)$ where $\zeta(t)$ is the Gaussian noise with $\langle\zeta(t)\zeta(t')\rangle=\delta(t-t')$, which gives rise to the terminal velocity $\langle\dot{x}(t)\rangle=F_{\text{ext}}/\gamma$. 
In contrast, the corresponding Langevin equation for an active particle is $\dot{x}(t)=V(u,w)+\sqrt{2D(u,w)}\zeta(t)$. In the latter case, both the velocity and diffusion constant at steady state are a function of substrate concentration, $u=u([\text{ATP}])$, the driving force of the particle's motion, which allows us to express $D$ as a function of $V$ such that $D=D(V)$.

To recapitulate, in this study we determined a set of microscopic rate constants, which best describe the ``trajectories'' of kinesin-1, on uni-cyclic kinetic model consisting of $N$-contiguous chemical states and transition rates between them, and evaluated the heat dissipation along the reaction cycle.  
The philosophy underlying the analysis of mapping trajectories on kinetic model, proposed here on kinesin-1 as well as others on F$_1$-ATPase \cite{Toyabe:2010bm,Shinagawa_2016_JPSJ}, is in essence similar to the one by the recent study which has quantified circulating flux on configurational phase space (or mode space) to diagnose broken DB and non-equilibrium dynamics at mesoscopic scale \cite{Battle2016Science,Gladrow2016PRL}. 
Lastly, our study confers quantitative insights into how much of the chemical free energy supplied to active systems (enzymes, molecular motors) is converted to mechanical movement in space and eventually dissipated into heat.
Variations in the transport properties and heat dissipation among different molecular motors provides glimpses into their design principles \cite{hinczewski2013PNAS2}, which should also be highlighted against typical enzymes specialized for catalysis.
\\

{\bf Analysis of kinesin-1 data using (N=4)-state kinetic model. }
The data digitized from ref. \cite{Visscher1999_Nature} were fitted to 
the ($N $=4)-state model used by Fisher and Kolomeisky \cite{Fisher01PNAS}, but we kept the parameter $w_4$ independent of [ATP]. 
Initial values for the fit were chosen from Eq.(14-15) in the Ref. \cite{Fisher01PNAS} except that we set $w_4 = 100$ $s^{-1}$ as an initial value for the fit.
The \emph{curve\_fit} from scipy \cite{Scipy} was used to globally fit the data in Fig.\ref{fig_VD}A-C, Fig.\ref{fig_kinesin_refit_fix}A-F.
$\theta_{4}^{-}$ is determined from the constraint $\sum_{n=1}^N (\theta_{n}^{+}+\theta_{n}^{-}) = 1$ \cite{Fisher99PNAS} at every iteration step.
The parameters determined from the fit shown in Fig.\ref{fig_kinesin_refit_fix} are provided in Table \ref{table}, and they are comparable to those in Ref. \cite{Fisher01PNAS}.

\section*{Aknowledgement}
We thank Bae-Yeun Ha, Hyunggyu Park, Dave Thirumalai, and Anatoly Kolomeisky for helpful comments and illuminating discussions.  
We acknowledge the Center for Advanced Computation in KIAS for providing computing resources.

\begin{table}[tbp]
	\caption{Parameters determined from the fit using (N=4)-state model. The unit of $\{u_n\}$ and $\{w_n\}$ is $s^{-1}$ except for $u_1^o$ ($[u_1^o]={\mu M}^{-1} s^{-1}$).}
	\begin{tabular}{ |c | l || c| l || c | l || c| l |  }
		\hline
		$u_1^0$ & $2.3$ & $u_2$ & $600$ & $u_3$ & $400$&$u_4$ & $190$ \\
		$\theta_{1}^{+}$ & $0.00$ & $\theta_{2}^{+}$ & $0.04$ & $\theta_{3}^{+}$ & $0.01$ &$\theta_{4}^{+}$ &$0.02$ \\
		$w_1$ & $20$  & $w_2$ & $1.4$ & $w_3$ & $1.7$ & $w_4$ & $120$ \\
		$\theta_{1}^{-}$ & $0.14$ &  $\theta_{2}^{-}$ & $0.15$& $\theta_{3}^{-}$ & $0.5$ &$\theta_{4}^{-}$ & $0.14$\\
		\hline
	\end{tabular}
	\label{table}
\end{table}
\FloatBarrier
%

\clearpage

%
\clearpage
\setcounter{figure}{0}  
\setcounter{equation}{0}
{\Large {\bf Supplementary Information} }

\renewcommand{\thefigure}{S\arabic{figure}} 
\renewcommand{\theequation}{S\arabic{equation}}
 \section{1. Derivation of the third degree polynomial dependence of $D$ on $V$.}
Here, we show a polynomial dependence of $D$ on $V$ using a few specific examples of the $N$-state periodic reaction model \cite{Derrida1983_JSP} whose reaction scheme is demonstrated in Fig.~\ref{fig_cyclic_model}.

\subsection{($N$=1)-state kinetic model}
When $N$=1, 
the master equation to solve is: 
\begin{align}
\dot{\pi}_\mu(t)=u_1\pi_{\mu-1}(t)+w_1\pi_{\mu+1}(t)-(u_1+w_1)\pi_\mu(t),
\label{eqn:master}
\end{align}
	where $\pi_\mu(t)$ is the probability of motor being in the $\mu$-th reaction cycle at time $t$.
Using generating function $F(z,t)=\sum_{\mu=-\infty}^{\infty}z^\mu \pi_\mu(t)$ with $\pi_\mu(0)=\delta_{\mu,0}$ \cite{vanKampen}, 
the master equation is written in terms of $F(z,t)$ as 
\begin{align}
\partial_tF(z,t)&=\left(u_1z+\frac{w_1}{z}-(u_1+w_1)\right)F(z,t)\nonumber\\
F(z,t)&=e^{\left(u_1z+\frac{w_1}{z}-(u_1+w_1)\right)t}.
\end{align}
Now, it is straightforward to obtain the mean velocity ($V$) and diffusion constant ($D$) using 
$\partial_z\log{F(z,t)}|_{z=1}=\langle \mu(t)\rangle$ and $\partial^2_z\log{F(z,t)}|_{z=1}=\langle \mu^2(t)\rangle-\langle \mu(t)\rangle^2-\langle \mu(t)\rangle$, where $\mu(t)$ is the number of steps taken by the molecular motor until time $t$. 
 \begin{equation} 
 V\equiv \lim_{t\rightarrow\infty}\frac{d_0\langle \mu(t)\rangle}{t}= d_0(u_1-w_1)
 \label{eq:1V}
 \end{equation}
 and
 \begin{equation}
 D\equiv \lim_{t\rightarrow\infty}\frac{d^2_0(\langle \mu^2(t)\rangle-\langle \mu(t)\rangle^2)}{2t} = \frac{d_0^2(u_1+w_1)}{2}.
 \label{eq:1D}
 \end{equation}
 Provided that only $u_1$ changes (for example by increasing ATP concentration) while $w_1$ remains constant. 
 elimination of $u_1$ from $V(u_1)$ and $D(u_1)$ relates $D$ to $V$ as 
 
 \begin{equation} \label{eq: D_V_1cycle}
 D(V) = D_0+\frac{d_0}{2}V
 \end{equation}
 where 
 \begin{align}
 D_0\equiv d_0^2w_1, 
 \label{eq:D0}
 \end{align}
 showing that for ($N$=1)-state kinetic model, $D$ is linear in $V$.

 \subsection{($N$=2)-state kinetic model} 
 For the ($N$=2)-kinetic model \cite{Fisher99PNAS}, 
 \begin{equation}
 V = d_0\frac{u_1 u_2 -  w_1 w_2}{u_1 + u_2 + w_1 + w_2}
 \label{eqn:V2}
 \end{equation}
 and
 \begin{equation}
 D=\frac{d_0^2}{2}\left[\frac{u_1u_2}{w_1w_2}+1-2\left(\frac{u_1u_2}{w_1w_2}-1\right)^2\frac{w_1w_2}{\sigma^2}\right]\frac{w_1w_2}{\sigma}
 \label{eqn:D2}
 \end{equation}
 where $\sigma=u_1+u_2+w_1+w_2$.  
 
 Then, $D=D(V)$ is obtained by eliminating $u_1$ between Eq.\ref{eqn:V2} and Eq.\ref{eqn:D2}:
 \begin{widetext}
 	\begin{align}
 	D(v)/d_0^2&=\left(\frac{u_2}{\kappa+1}\right)+\frac{1}{2}\left(\frac{\kappa-1}{\kappa+1}\right)u_2v-\left(\frac{u_2}{\kappa+1}\right)\frac{u_2^2}{w_1w_2}v^2+\left(\frac{u_2}{\kappa+1}\right)\frac{u_2^2}{w_1w_2}v^3
	\label{eq:D_V_2cycle}
 	\end{align}
 \end{widetext}
 where $\kappa\equiv\frac{u_2(u_2+w_1+w_2)}{w_1w_2}$, and $v\equiv V/V_{max}$ ($0\leq v\leq 1$) with $V_{max}=d_0u_2$.  
 Eq.\ref{eq:D_V_2cycle} confirms that $D$ is a third order polynomial in $V$.  

 Incidentally, the ($N$=2)-kinetic model is reduced to the Michaelis-Menten equation by setting $u_1=u_1^o[S]$ and $w_2=0$. 
 \begin{equation}
 V=  d_0\frac{u_1 u_2}{u_1 + u_2 + w_1} = \frac{V_{max}[S]}{ K_M + [S] } 
 \end{equation}
 where $K_M = (u_2 + w_1)/u_1^o$, and 
 \begin{align} 
 D(v)
 =D_{max}v\left[1-2\phi v+2\phi v^2\right]
 \label{eq:V_D_MM}
 \end{align}
 with $\phi\equiv \frac{k_{cat}}{u_1^oK_M}$. 
 Note that $D\leq D_{max}=d_0V_{max}/2$ and that for $D(v)$ to be positive for all the range of $v$, 
 the parameter $\phi$ should be in a rather narrow range of $0\leq \phi \leq 2$.
 
 \subsection{General case: $N$-state kinetic model}
 The above two examples of one-dimensional hopping model was extended to the $N$-state kinetic model by Derrida  \cite{Derrida1983_JSP}. 
 He obtained exact expressions for the mean velocity ($V^{(D)}$) and diffusion constant ($D^{(D)}$), where the superscript $(D)$ refers to Derrida's, in terms of the rate constants $\{u_n\}$ and $\{w_n\}$. 
 Derrida's expression for $V^{(D)}$ and $D^{(D)}$ are related to $V$ and $D$ as $V\equiv (d_0/N)V^{(D)}$  and $D\equiv (d_0^2/N^2)D^{(D)}$. 
 \begin{align} 
 V^{(D)} = \frac{N} {\sum_{n=1}^{N} r_n}\left(1- \frac{\prod_{n=1}^{N} w_n}{\prod_{n=1}^{N} u_n }\right)
 \label{eq: V_Derrida}
 \end{align}
 and  
 \begin{widetext}
 	\begin{align}\label{eq:D_Derrida}
 	D^{(D)}&=\frac{1}{\left(\sum_{n=1}^Nr_n\right)^2}\left[V^{(D)}\sum_{n=1}^Nq_n\sum_{i=1}^Nir_{n+i}+N\sum_{n=1}^Nu_nq_nr_n\right]
	-V^{(D)}\frac{N+2}{2}
 	\end{align}
 \end{widetext}
 where 
 $r_n = \frac{1}{u_n} \left[1 + \sum_{i=1}^{N-1} \prod_{j=1}^{i} \frac{w_{n+j-1}}{u_{n+j}}\right]
 $,
  and 
 $q_n=\frac{1}{u_n}\left[1+\sum_{i=1}^{N-1}\prod_{j=1}^i\frac{w_{n-j}}{u_{n-j}}\right]
 $
  with periodic boundary conditions $u_{n+N} = u_n$, and $w_{n+N} = w_n$.
  $D^{(D)} = D^{(D)}(V^{(D)})$ is obtained by eliminating $u_1$ between Eq.\ref{eq: V_Derrida} and Eq.\ref{eq:D_Derrida}.
  For that, we first express various terms in Eq.\ref{eq:D_Derrida} in terms of $u_1$:  
 \begin{align} \label{eq:r_ABu1}
 \sum_{n=1}^Nr_n&=\frac{A}{u_1}+B,\nonumber\\
 \sum_{n=1}^Nq_n\sum_{i=1}^Nir_{n+i}&=\frac{\alpha}{u_1^2}+\frac{\beta}{u_1}+\gamma,\nonumber\\
  \sum_{n=1}^Nu_nq_nr_n&=\frac{\xi}{u_1^2}+\frac{\eta}{u_1}+\zeta.
  \end{align}
 where A, $B\left(=\sum_{n=2}^N\frac{1}{u_n}\left[1+\sum_{i=1}^{N-n}\prod_{j=1}^i\frac{w_{n+j-1}}{u_{n+j}}\right]\right)$,
 $\alpha$, $\beta$, $\gamma$, $\xi$, $\eta$, and $\zeta$ are all positive constants independent of $u_1$. 
 Next, $\sum_{n=1}^Nr_n$ in Eq.\ref{eq:r_ABu1} substituted to Eq.\ref{eq: V_Derrida} gives 
 \begin{align}  \label{eq:V(u_1)} 
 V^{(D)}=N\frac{1-C/u_1}{A/u_1+B},
 \end{align}
 where 
 $C\left(= \prod_{n=1}^Nw_n/\prod_{n=2}^Nu_n \right)$, and $u_1$ is expressed in terms of $V^{(D)}$
 \begin{align}
\frac{1}{u_1}=\frac{1-B\frac{V^{(D)}}{N}}{C+A\frac{V^{(D)}}{N}}. 
\label{eq:u_1(V)}
 \end{align}
 Finally, with Eqs.\ref{eq:r_ABu1} and 
 \ref{eq:u_1(V)}, we show that $D^{(D)}$ (Eq.\ref{eq:D_Derrida}) can be expressed as a third degree polynomial in $V^{(D)}$ 
 \begin{widetext}
 	\begin{align}
 	D^{(D)}
 	&=V^{(D)}\frac{\left[\alpha(1-B\frac{V^{(D)}}{N})^2+\beta(1-B\frac{V^{(D)}}{N})(A\frac{V^{(D)}}{N}+C)+\gamma(A\frac{V^{(D)}}{N}+C)^2\right]}{(A+BC)^2}\nonumber\\
 	&\hspace{0.5cm}+N\frac{\xi(1-B\frac{V^{(D)}}{N})^2+\eta(1-B\frac{V^{(D)}}{N})(A\frac{V^{(D)}}{N}+C)+\zeta(A\frac{V^{(D)}}{N}+C)^2}{(A+BC)^2}-V^{(D)}\frac{N+2}{2}\nonumber\\
 	&=z_0+z_1V^{(D)}+z_2(V^{(D)})^2+z_3(V^{(D)})^3\nonumber\\
 	D&=\alpha_0+\alpha_1V+\alpha_2V^2+\alpha_3V^3. 
 	\label{eqn:DV}
  	\end{align}
 
 	with $\alpha_i=(d_0/N)^{2-i}z_i$, and 
 	\begin{align}
 	z_0&=\frac{(\xi+\eta C+\zeta C^2)}{(A+BC)^2}\nonumber\\
 	z_1&=\left[\frac{(\alpha+\beta C+\gamma C^2)+(-2\xi B+\eta (A-BC)+2\zeta AC)}{(A+BC)^2}-\frac{N+2}{2}\right]\nonumber\\
 	z_2&=\frac{(-2\alpha B+\beta (A-BC) +2\gamma AC)+(\xi B^2-\eta AB +\zeta A^2)}{(A+BC)^2}\nonumber\\
 	z_3&=\frac{(\alpha B^2-\beta AB +\gamma A^2)}{(A+BC)^2}. \nonumber
 	\end{align}
 \end{widetext}
\clearpage 

 \subsection{Alternative derivation of $D(V)$} 
In addition to Derrida's result \cite{Derrida1983_JSP}, the sign of $\alpha_i$ can be determined by deriving the relation between $V^{ (D) }$ and $D^{ (D) }$ using the result in ref. \cite{Koza1999}.
 From Eq.(23) in ref. \cite{Koza1999},
 \begin{equation} \label{eq: V_Koza}
 \begin{aligned}
 V^{ (D) } = - i \frac{c'_0}{c_1}
 \end{aligned}
 \end{equation}
where $c'_0 = i N ( \prod_{n=1}^{N} u_n - \prod_{n=1}^{N} w_n )$ and $c_1 = c_1( \{u_n\}, \{w_n\})$. 
By combining two expressions of $V^{ (D) }$, Eq.\ref{eq: V_Derrida} and Eq.\ref{eq: V_Koza}, we get
\begin{widetext}
\begin{align}
c_1&  = \prod_{n=1}^{N} u_n\times \sum_{m=1}^{N-1} \left[ \frac{1}{u_m} \left( 1 + \sum_{i=1}^{N-1} \prod_{j=1}^{i} \frac{w_{m+j}}{u_{m+j}}\right) \right] \nonumber\\
 &  = u_1 \prod_{n=2}^{N} u_n \left[\frac{1}{u_1} \left( 1 + \sum_{i=1}^{N-1} \prod_{j=1}^{i} \frac{w_{1+j}}{u_{1+j}} \right)
 + \sum_{m=2}^{N-1} \frac{1}{u_m} \left( 1 + \sum_{i=1}^{N-1} \prod_{j=1}^{i} \frac{w_{m+j}}{u_{m+j}} \right) 
 \right]\nonumber\\
 &  = \mathcal{A}  u_1 + \mathcal{B}
 \label{eq: c_1}
 \end{align}
 \end{widetext}
 \noindent where $\mathcal{A}$ and $\mathcal{B}$ are constants depending on ($u_2,\ldots,u_N$) and ($w_1,w_2,\ldots w_N$). 
 Eq.(\ref{eq: c_1}) substituted to Eq.(\ref{eq: V_Koza}) gives 
 \begin{equation}
 \begin{aligned}
 V^{ (D) } = N \frac{u_1 \prod_{n=2}^{N} u_n - \prod_{n=1}^{N} w_n}{\mathcal{A}u_1 + \mathcal{B}}
 \end{aligned}
 \end{equation}
 and hence $u_1$ can be written as
 \begin{equation}\label{eq:u_1(V)_2}
 \begin{aligned}
 u_1 = \frac{\mathcal{B}  V^{ (D) } + N \prod_{n=1}^N w_n}{N \prod_{n=2}^{N} u_n - \mathcal{A}V^{ (D) }}.
 \end{aligned}
 \end{equation}
 From Eq.\ref{eq:u_1(V)} and Eq.\ref{eq:u_1(V)_2}, 
 $\mathcal{A} = B \prod_{n=2}^N u_n$ and $\mathcal{B} = A \prod_{n=2}^N u_n$ where $A$ and $B$ are the same constants used in Eq.\ref{eq:V(u_1)}.
 Now using Eq.(\ref{eq: c_1}), Eq.(\ref{eq:u_1(V)}), and general expression of $D^{ (D) }$ from Eq.(24) of ref. \cite{Koza1999}, we have
 \begin{equation}
 \begin{aligned}
 D^{ (D) } & = \frac{c_0'' - 2 c_2 (V^{ (D) })^2 }{2 c_1 } \\
  &=\frac{c_0'' - 2 c_2 (V^{ (D) })^2 }{2 (\frac{B}{\prod_{n=2}^N u_n}  u_1 + \frac{A}{\prod_{n=2}^N u_n}) } \\
  & = \frac{( c_0'' - 2 c_2 (V^{ (D) })^2 ) (\prod_{n=2}^{N} u_n (N- B V^{(D)}) )}{2N( A + C B)}
 \end{aligned}
 \end{equation}	
 where $c_0'' = N^2 ( u_1 \prod_{n=2}^{N} u_n + \prod_{n=1}^{N} w_n ) = N^2 (\prod_{n=2}^N u_n) (u_1 + C)$, and $c_2 = \beta_1 u_1 + \beta_2$ where $\beta_1$ and $\beta_2$ 
 are positive constants depending on ($u_2, u_3, \cdots, u_N$) and ($w_1, w_2, \cdots, w_N$) (see Eq.(53, 54) of ref. \cite{Koza1999}).
 Since $u_1 (N - B V^{ (D) })  = NC + AV^{ (D) }$ (Eq.\ref{eq:V(u_1)}), we get
 \begin{widetext}
 \begin{align}
 D^{ (D) } & = \frac{1}{2} \frac{( c_0^{''} - 2 c_2 (V^{ (D) })^2 ) ( (N- B V^{(D)}) )\prod_{n=2}^{N} u_n}
 {N A + N C B} \nonumber\\ 
 & = \frac{1}{2} \frac{( N^2 (\prod_{n=2}^N u_2) (u_1 + C) - 2 (\beta_1 u_1 + \beta_2) (V^{ (D) })^2 ) ( (N- B V^{(D)}) )\prod_{n=2}^{N} u_n}
 {N A + N C B} \nonumber\\ 
 & = \frac{\prod_{n=2}^N u_n}{2} \frac{ \left( N^2 (\prod_{n=2}^N u_2) ( NC + A V^{(D)} + C(N-B V^{ (D) } ) ) - 2 ( \beta_1 (N C + A V^{(D)}) + \beta_2 ( N- B V^{(D)}  ) ) (V^{ (D) })^2 \right)}
 {N A + N C B} \nonumber\\
  &= z_0 + z_1 V^{ (D) } + z_2 (V^{ (D) })^2 + z_3 (V^{ (D) })^3
 \label{eq:D_tr2}
 \end{align}
  \end{widetext}		
 where 
 \begin{align}
 z_0&=\frac{N^2 (\prod_{n=2}^N u_n )^2 C}{ A + BC }>0\nonumber\\
 z_1&=\frac{N (\prod_{n=2}^N u_n )^2  (A- BC)}{2 (A + BC)}\nonumber\\
 z_2&=- \frac{(\prod_{n=2}^N u_n )( \beta_1 C + \beta_2 )}{A + BC}<0\nonumber\\
 z_3&= \frac{(\prod_{n=2}^N u_n )( - \beta_1 A + \beta_2 B)}{N(A + BC)}.\nonumber
 \end{align}
It is obvious that $z_0 >0$ and $z_2 < 0$ since $A$, $B$, $C$, $\beta_1$, and $\beta_2$ are all positive constants.  

 
  \section{2. The 1D hopping model with a finite processivity}
Because of a probability of being dissociated from microtubules, kinesin motors display a finite processivity.   
 However, since the mean velocity and diffusion constant are calculated from the trajectories that remain on the track, 
 the expressions of $V$ and $D$ in terms of the rate constants are unchanged. 
To make this point mathematically more explicit, we consider the master equation assuming a constant dissociation rate $k_d$ from each chemical state.
   \begin{align}\label{eq:masterequation_kd}
   \frac{d P_{\mu,n} (t)}{d t}&=u_{n-1} P_{\mu, n-1}(t)+w_{n}P_{\mu, n+1}(t)\nonumber\\
   &-(u_n+w_{n-1}+k_d)P_{\mu, n}(t)
   \end{align}
   where $P_{\mu, n}(t)$ is the probability of being in the $n$-th chemical state at the $\mu$-th reaction cycle.
   The probability of the motor remaining on the track (survival probability of motor) is  
   \begin{align} \label{eq:P_surv}
   S (t) \equiv \sum_{\mu=\infty}^{\infty} \sum_{n=1}^{N} P_{\mu, n}(t) = e^{-k_d t}.
   \end{align}
   The expectation value of an observable, which can be used to calculate $\langle x(t)\rangle$ or $\langle x^2(t)\rangle$,  is expressed as
   \begin{align}
   \langle A(t)\rangle = \sum_{\mu=-\infty}^{\infty} \sum_{n=1}^{N} \Phi_{\mu,n}(t) A(\mu(t))
   \end{align}
   with a probability density function renormalized with respect to the survival probability 
   \begin{align} \label{eq:f_mu,n}
   \Phi_{\mu, n}(t) \equiv \frac{P_{\mu,n}(t)}{S(t)}=P_{\mu,n}(t)e^{k_dt}. 
   \end{align}
   Incidentally, $\Phi_{\mu, n}(t)$ satisfies the following master equation. 
   \begin{align}
   \frac{d\Phi_{\mu,n} (t)}{dt}&=u_{n-1}\Phi_{\mu, n-1}(t)+w_{n}\Phi_{\mu, n+1}(t) \\
   &\hspace{1cm}-(u_n+w_{n-1})\Phi_{\mu, n}(t),
 \end{align}
 which is identical to Eq.\ref{eq:masterequation}, but now the probability of interest is explicitly confined to the ensemble of trajectories  remaining on the track. 
   %
  For an arbitrary value of $k_d$ and for any $N$, the expressions of $V$, $D$, and Eq.\ref{eqn:DV} remain unchanged except that the range of ensemble is specific to the motor trajectories remaining on the track. 
   Furthermore, the expression of $\dot{Q}$, which depends only on $V$ and rate constants, remains identical in the presence of detachment (finite $k_d>0$). 
   Therefore, our formalisms remain valid for motors with a finite processivity.

 \section{3. Mapping the master equation for $N$-state kinetic model onto Langevin and Fokker-Planck equations}
 The master equation (Eq.\ref{eqn:master}) can be mapped onto a Langevin equation for position $x(t)$ as 
 \begin{align}
\dot{x}(t)=V+\sqrt{2D}\eta(t)
\label{eq:Langevin}
 \end{align}
 where for (N=1)-state model $V=d_0(u_1-w_1)$ and $D=d_0^2/2\times (u_1+w_1)$ as in Eqs.\ref{eq:1V} and \ref{eq:1D}, and 
 $P[\eta(t)]\propto \exp{\left(-\frac{1}{2}\int_0^td\tau \eta^2(\tau)\right)}$. 
Then, with the transition probability (propagator), 
\begin{align}
P(x_{t+\epsilon}|x_t)&=\left(\frac{1}{4\pi D\epsilon}\right)^{1/2}e^{-\frac{\{x_{t+\epsilon}-x_t-\epsilon V\}^2}{4D\epsilon}}, 
\end{align}
where $x_{t}\equiv x(t)$,  
and starting from an initial condition, $P[x(0)]=\delta[x(0)-x_0]$, it is straightforward to obtain  
the position of motor at time $t$: 
\begin{widetext}
\begin{align}
P[x(t)]&=\int dx_0\int dx_{\epsilon}\cdots \int dx_{t-\epsilon}P(x_t|x_{t-\epsilon})\cdots P(x_{2\epsilon}|x_{\epsilon})P(x_{\epsilon}|x_0)P(x_0) \nonumber\\
&=\left(\frac{1}{4\pi Dt}\right)^{1/2}\exp{\left(-\frac{(x(t)-x_0-Vt)^2}{4Dt}\right)}\nonumber\\
  &=\left(\frac{1}{2\pi d_0^2(u_1+w_1)t}\right)^{1/2} \exp{\left(-\frac{[x(t)-x_0-d_0(u_1-w_1)t]^2}{2d_0^2(u_1+w_1)t}\right)},  
 \end{align}
 \end{widetext}
\noindent where we plugged $V$ and $D$ from Eqs.\ref{eq:1V}, \ref{eq:1D} for ($N$=1)-state kinetic model in the last line.  
Unlike the normal Langevin equation, where the noise strength determined by FDT is associated with an ambient temperature ($\sim\sqrt{T}$),  
the noise strength in Eq.\ref{eq:Langevin} is solely determined by the forward and backward rate constants, which fundamentally differs from the Brownian motion of a thermally equilibrated colloidal particle in a heat bath.

Next, Fokker-Planck equation follows from Eq.\ref{eq:Langevin}, 
\begin{align}
\partial_t P(x,t) &= D \partial_x^2 P(x,t) - V \partial_x P(x,t)\nonumber\\
&=-\partial_x j(x,t)
\label{eq:Fokker}
\end{align}
with the probability current being defined as
\begin{align}
j (x,t) = - D \partial_x P(x,t) + V P(x,t).
\label{eq:flux}
\end{align}
Then, \emph{mean local velocity} $v(x,t)\equiv j(x,t)/P(x,t)$ is defined
\begin{align}
v(x,t)= V - D  \partial_x \log{ P(x,t) }. 
\label{eq:localvelocity}
\end{align}
In order to relate this definition of the mean local velocity to heat dissipated from the molecular motor moving along microtubules in a NESS, 
we consider $\gamma_{\text{eff}}$, \emph{an effective friction coefficient}, and introduce a nonequilibrium potential $\phi(x)\equiv-\log{P^{ss}(x)}$ \cite{Seifert2012RPP,Hatano2001_PRL}. 
By integrating the both side of Eq.\ref{eq:localvelocity} in a NESS with respect to the displacement corresponding to a single step, we obtain  
\begin{align}
\int_x^{x+d_0} \gamma_{\text{eff}} v^{ss}(x)dx &= \gamma_{\text{eff}} V d_0 +\gamma_{\text{eff}}D(\phi(x+d_0)-\phi(x)).  
\label{heat}
\end{align}
Following the literature on NESS thermodynamics \cite{Hatano2001_PRL,Qian_2006_JPCB,Seifert2012RPP}, we endowed each term of Eq.\ref{heat} with its physical meaning.   
(i) housekeeping heat:
\begin{align}
Q_{hk}=\int_x^{x+d_0} \gamma_{\text{eff}} v(x)dx,
\label{eq:Qhk} 
\end{align}
(ii) total heat: 
\begin{align}
Q=\gamma_{\text{eff}} V d_0
\label{eq:Qeff}
\end{align}
and (iii) excess heat:  
\begin{align}
Q_{ex}=-\gamma_{\text{eff}}D(\phi(x+d_0)-\phi(x)). 
\label{eq:Qex}
\end{align}
Eqs.\ref{eq:Qhk}, \ref{eq:Qeff}, and \ref{eq:Qex} satisfy 
\begin{align}
Q_{hk}=Q-Q_{ex}. 
\end{align}
and in fact $Q_{ex}=0$ because of the periodic boundary condition implicit to our problem of molecular motor, which leads to $\phi(x+d_0)=\phi(x)$. 
Hence,  
\begin{align}
Q_{hk}=Q=\gamma_{\text{eff}}Vd_0. 
\label{eq:heatNESS}
\end{align}
Although we introduced the effective friction coefficient $\gamma_{\text{eff}}$ in Eq.\ref{heat} to define the heats produced at nonequilibrium, Eq.\ref{eq:heatNESS} finally allows us to associate $\gamma_{\text{eff}}$ with other physically well-defined quantities. 
\begin{align}
\gamma_{\text{eff}}=\frac{Q_{hk}}{Vd_0} = \frac{k_B T}{d^2_0(j_+-j_-)} \log{ \left( \frac{j_+}{j_-} \right) }.
\label{eq:gammaeff}
\end{align}
Here, note that we for the first time introduced the temperature $T$, which was discussed neither in the master equation (Eq.\ref{eqn:master}) nor in the Langevin equation (Eq.\ref{eq:Langevin}). 
Of special note is that $\gamma_{\text{eff}}$ does not remain constant, but depends on the steady state flux $j^{ss}=j=j_+-j_-$. 
Similar to the effective diffusion constant of bacterium, $D_{\text{eff}}$, discussed in the main text, $\gamma_{\text{eff}}$ is defined operationally.  
At equilibrium, when the detailed balance (DB) is established ($j_+=j_-=j_0$), $\gamma_{\text{eff}}^{\text{DB}}$ approaches to: 
\begin{align}
\gamma_{\text{eff}}^{\text{DB}} &=\lim_{j_+\rightarrow j_-}\frac{k_B T}{d^2_0(j_+-j_-)} \log{ \left( \frac{j_+}{j_-} \right) }\rightarrow \frac{k_B T}{d^2_0j_{0}}.  
\end{align} 
In fact, 
$D_0=k_BT/\gamma_{\text{eff}}^{\text{DB}}=d_0^2j_0$ satisfies the FDT for passive particle at thermal equilibrium, i.e., $k_BT=D_0\gamma^{\text{DB}}_{\text{eff}}$. 
For ($N$=1)-state model, $D_0=d_0^2w_1$, which is identical to Eq.\ref{eq:D0}.  \\

 \section{4. Nonequilibrium steady state thermodynamics.}
 To drive a system out of equilibrium, one has to supply a proper form of energy into the system. 
 Molecular motors move in one direction because transduction of chemical free energy into conformational change is processed.  
 Relaxation from a nonequilibrium state is accompanied with heat and entropy production. 
 In the presence of external nonconservative force (chemical or mechanical force), the system reaches the nonequilibrium steady state. 
 If one considers a Markov dynamics for microscopic state $i$, 
 described by the master equation $\partial_t p_i(t)=-\sum_j (W_{ij}p_i(t)-W_{ji}p_j(t))$, 
 the system relaxes to nonequilibrium steady state at long time, establishing time-independent steady state probability $\{p_i^{ss}\}$ for each state satisfying the zero flux condition $\sum_j(W_{ij}p^{ss}_i-W_{ji}p^{ss}_j)=0$.  
 A removal of the nonconservative force is led to further relaxation to the equilibrium ensemble, in which the detailed balance (DB) is (locally) established in every pair of the states such that $p_i^{eq}W_{ij}=p_j^{eq}W_{ji}$ for all $i$ and $j$.  
 An important feature of the equilibrium, which differentiates itself from NESS, is the condition of DB.

 Over the decade, there have been a number of endeavors to better characterize the system out-of-equilibrium \cite{Seifert2012RPP}. 
 One of them is to define the heat and entropy production in the context of Master equation. 
 The heat and entropy productions in reference to either steady state or equilibrium are defined to better characterize the process of interest.
 The aim is to associate the time dependent probability for state ($\{p_i(t)\}$) and transition rates between the states $\{W_{ij}\}$ with newly defined macroscopic thermodynamic quantities at nonequilibrium \cite{Qian2010PRE}. 
 Here, we review NESS thermodynamics formalism developed by Ge and Qian \cite{Qian2010PRE}.

 For nonequilibrium relaxation processes one can consider three  relaxation processes: 
 (i) relaxation process of a system far-from-equilibrium (FFE) to a nonequilibrium steady state (NESS); 
 (ii) relaxation process of a system far-from-equilibrium (FFE) to an equilibrium (EQ). 
 (iii) relaxation process of a system in NESS to an equilibrium (EQ).  
 To describe these relaxation processes using the probabilities for state, we introduce a phenomenological definition of an internal energy of state $i$ at a steady state by $u^{ss}_i=-k_BT\log{p_i^{ss}}$, and at equilibrium by $u^{eq}_i=-k_BT\log{p_i^{eq}}$.  
 Then the following thermodynamic quantities are defined either in reference to NESS or equilibrium.  
 
 First, the thermodynamic potentials are defined in reference to the NESS: 
 the total energy $U(t)=\sum_i^Np_i(t)u^{ss}_i$; 
 the total free energy $F(t)=U(t)-TS(t)=k_BT\sum_i^Np_i(t)\log{(p_i(t)/p_i^{ss})}$. 
 Second, the thermodynamic potentials are defined in reference to the equilibrium: 
 the total energy $U^{eq}(t)=\sum_i^Np_i(t)u^{eq}_i$; 
 the total free energy $F^{eq}(t)=U^{eq}(t)-TS(t)=k_BT\sum_i^Np_i(t)\log{(p_i(t)/p_i^{eq})}$. In both cases, Gibbs entropy, $S(t)=-k_B\sum_i^Np_i(t)\log{p_i(t)}$, is defined as usual.

\begin{figure}[ht]
 	\centering 
 	\includegraphics[scale=0.27]{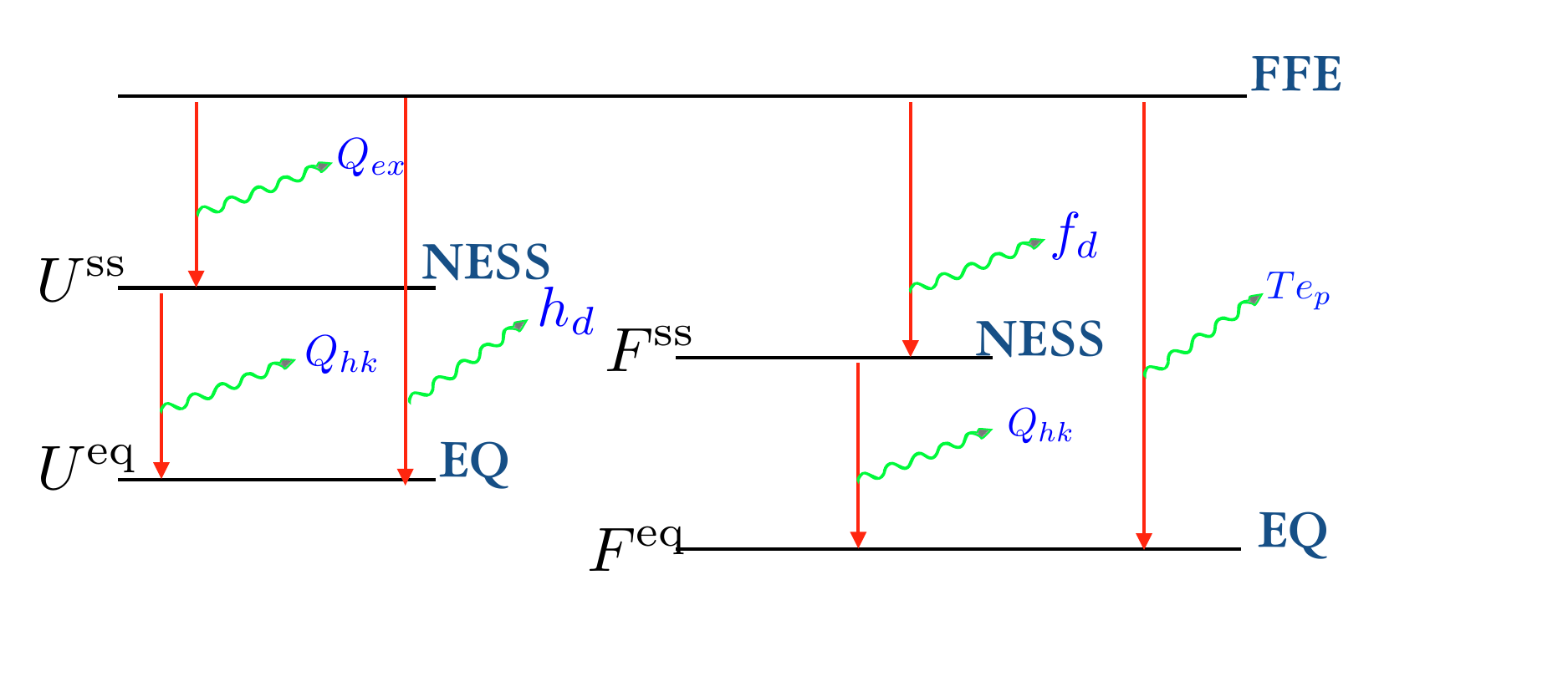}
 	\caption{A diagram illustrating the balances between various thermodynamic quantities discussed in the text. The curvy arrows denote the heat and entropy production from relaxation processes. }
 	\label{diagram}
 \end{figure}

 Next, the above definitions of generalized thermodynamic potentials, 
 one can define the heat and entropy productions associated with the relaxation processes (i), (ii), (iii). 
 The diagram in Fig.\ref{diagram} depicts the relaxation processes mentioned here. 
 
 \begin{widetext}
  $-dF(t)/dt$ is the rate of entropy production in the relaxation from FFE to NESS, 
 	\begin{align}
 	\frac{dF(t)}{dt}\equiv-\dot{f}_d&=-T\sum_{i>j}\left[W_{ij}p_i(t)-W_{ji}p_j(t)\right]\log{\left[\frac{p_i(t)p_j^{ss}}{p_j(t)p_i^{ss}}\right]}. 
 	\label{fd}
 	\end{align}
 	and $-dU(t)/dt$ is the rate of heat production in the relaxation from FFE to NESS. 
 	\begin{align}
 	\frac{dU(t)}{dt}\equiv-\dot{Q}_{ex}&=-\sum_{i>j}(W_{ij}p_i(t)-W_{ji}p_j(t))(u^{ss}_i-u^{ss}_j)\nonumber\\
 	&=T\sum_{i>j}(W_{ij}p_i-W_{ji}p_j)\log{\left(\frac{p_i^{ss}}{p_j^{ss}}\right)}. 
 	\label{Qex}
 	\renewcommand{\theequation}{S\arabic{align}}
 	\end{align}
 	Similarly, $-dF^{eq}(t)/dt$ is the rate of entropy production during the relaxation to equilibrium, 
 	\begin{align}
 	\frac{dF^{eq}(t)}{dt}\equiv-T\dot{e}_p&=-T\sum_{i>j}\left[W_{ij}p_i(t)-W_{ji}p_j(t)\right]\log{\left[\frac{p_i(t)W_{ij}}{p_j(t)W_{ji}}\right]}
 	\label{Tep}
 	\end{align}
 	and $-dU^{eq}(t)/dt$ is the rate of heat production.
 	\begin{align}
 	\frac{dU^{eq}(t)}{dt}\equiv-\dot{h}_d&=-\sum_{i>j}(W_{ij}p_i(t)-W_{ji}p_j(t))(u^{eq}_i-u^{eq}_j)\nonumber\\
 	&=T\sum_{i>j}(W_{ij}p_i-W_{ji}p_j)
 	\log{\left(\frac{W_{ij}}{W_{ji}}\right)}
 	\label{hd}
 	\end{align}
 \end{widetext}
 where the condition of DB ($p_i^{eq}/p_j^{eq}=W_{ji}/W_{ij}$) was used to derive the last line. 
 Furthermore, the heat production involved with the relaxation from NESS to equilibrium ($\dot{Q}_{hk}$), namely housekeeping heat which is introduced in NESS stochastic thermodynamics from the realization that maintaining NESS requires some energy, is defined by either using $\dot{Q}_{hk}=(-dF^{eq}(t)/dt)-(-dF(t)/dt)=T\dot{e}_p-\dot{f}_d$ or $\dot{Q}_{hk}=(-dU^{eq}(t)/dt)-(-dU(t)/dt)=\dot{h}_d-\dot{Q}_{ex}$. 
 Explicit calculations using the representation of thermodynamic potential in terms of master equation lead to 
 \begin{align}
 \dot{Q}_{hk}=T\sum_{i>j}\left[W_{ij}p_i(t)-W_{ji}p_j(t)\right]\log{\left[\frac{p^{ss}_iW_{ij}}{p_j^{ss}W_{ji}}\right]}. 
 \label{Qhk}
 \end{align}
 Lastly, from the definition of Gibbs entropy ($S(t)=-k_B\sum_ip_i(t)\log{p_i(t)}$), or from the thermodynamic relationships $TdS/dt=dF/dt-dU/dt=dF^{eq}/dt-dU^{eq}/dt$, it is straightforward to show that 
 \begin{align}
 T\frac{dS}{dt}&=\frac{dF}{dt}-\frac{dU}{dt}=\frac{dF^{eq}}{dt}-\frac{dU^{eq}}{dt}\nonumber\\
 &=-T\sum_{i>j}\left[W_{ij}p_i(t)-W_{ji}p_j(t)\right]\log{\left(\frac{p_i(t)}{p_j(t)}\right)}\nonumber\\
 &=\dot{h}_d-T\dot{e}_p.   
 \label{dSdt}
 \end{align}
 
 Now, with the various heat and entropy production defined from generalized potentials $F(t)$, $U(t)$, and $F^{eq}(t)$, $U^{eq}(t)$ ($\dot{f}_d$, $\dot{Q}_{ex}$, $T\dot{e}_p$, $\dot{h}_d$, and $\dot{Q}_{hk}$) we acquire two important balance laws in nonequilibrium thermodynamics: 
 \begin{align}
 T\dot{e}_p&=\dot{f}_d+\dot{Q}_{hk}\nonumber\\
 \dot{h}_d&=\dot{Q}_{hk}+\dot{Q}_{ex}
 \end{align} 
 Thus, (i) the total entropy production of a system, $T\dot{e}_p(=-dF^{eq}/dt)$, is contributed by the free energy dissipation due to the relaxation to NESS, $\dot{f}_d(=-dF/dt)$, and the housekeeping heat, $\dot{Q}_{hk}(=dF/dt-dF^{eq}/dt)$, that is required to maintain the NESS. (ii) The total heat production $\dot{h}_d(=-dU^{eq}/dt)$ of a system is decomposed into $\dot{Q}_{hk}(=dU/dt-dU^{eq}/dt)$ and the excess heat $\dot{Q}_{ex}(=-dU/dt)$.  
 The diagram in Fig.\ref{diagram} recapitulates the various heat and entropy production terms and their balance. 
 When the system is already in NESS, then neither the production of entropy nor excess heat is anticipated ($\dot{f}_d=0$, $\dot{Q}_{ex}=0$), and hence it follows that the amount of heat, entropy, and housekeeping heat required to sustain NESS are identical ($T\dot{e}_p=\dot{Q}_{hk}=\dot{h}_d$).
 
  \begin{figure*}[ht]
  	\centering 
  	\includegraphics[scale=0.5]{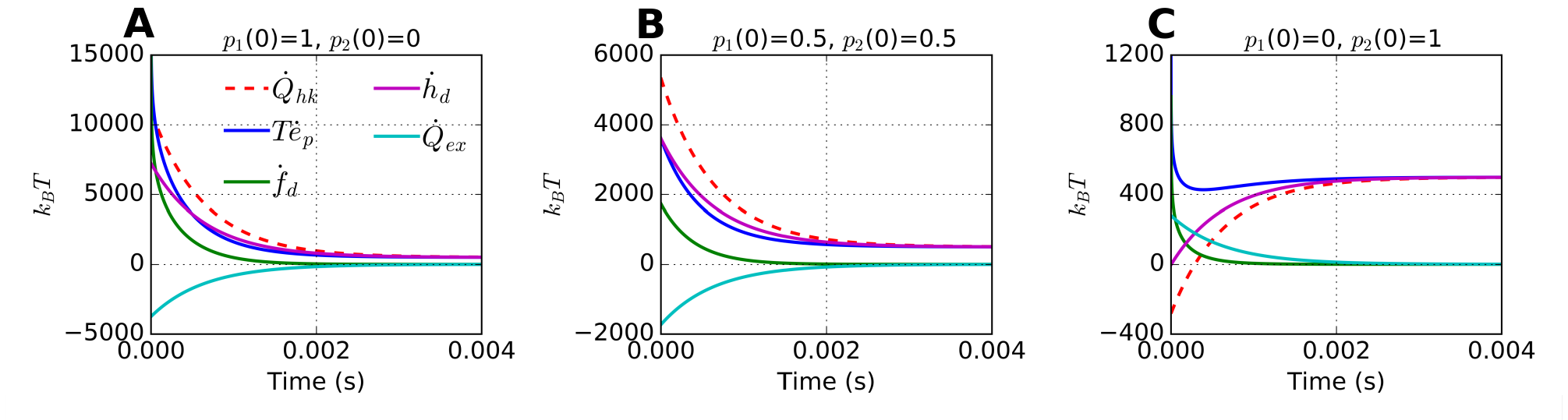}
  	\caption{Relaxation dynamics of various nonequilibrium thermodynamic quantities from far-from-equilibrium states calculated using (N=2)-state system.
  		The parameters used for the plots are: 
  		[ATP] = 1 mM, $f$ = 1 pN; 
  		$u_1^0 = 1.8 ~ s^{-1} \mu M^{-1}$, $u_2 = 108 ~ s^{-1}$, $w_1 = 6.0 ~ s^{-1}$, and $w_2 = 16 ~ s^{-1}$ at zero load;
  		$\theta^+_1 = 0.135, \theta^+_2 = 0.035, \theta^-_1 = 0.080$, and $\theta^-_2 = 0.75$.  
  		Plots were made using three different initial conditions: {\bf A.} $p_1(0) = 1, p_2(0) = 0$; {\bf B.} $p_1(0) = 0.5, p_2(0) = 0.5$; and {\bf C.} $p_1(0)=0, p_2(0) =1$.
  	}
  	\label{fig_2states_relax}
  \end{figure*}  
  
 In order to gain a better insight into the energy balance of molecular motor that operates in nonequilibrium steady state, we consider the dynamics of molecular motor systems by means of a cyclic Markov model and relate the essential parameters of the model with NESS thermodynamics.  
 The thermodynamic quantities associated with nonequilibrium process ($\dot{f}_d$, $\dot{Q}_{hk}$, $T\dot{e}_p$, $\dot{h}_d$, $\dot{Q}_{ex}$) can be evaluated explicitly using $(N=2)$-state Markov model; 
 the time evolution of each state is given by 
 $p_1(t)=p_1^{ss}+(p_1(0)-p_1^{ss})e^{-\sigma t}$ and $p_2(t)=p_2^{ss}-(p_1(0)-p_1^{ss})e^{-\sigma t}$
 with $p_1^{ss}=(u_2+w_1)/\sigma$, $p_2^{ss}=(u_1+w_2)/\sigma$, and $\sigma=u_1+u_2+w_1+w_2$. 
 Using the conditions satisfied in 2-state model ($p_{1+2}(t)=p_1(t)$) $W_{12}=u_1$, $W_{21}=w_1$, $W_{23}=u_2$, $W_{32}=w_2$; otherwise $W_{ij}=0$, we obtain 
 \begin{widetext}
 	\begin{align}
 	\frac{\dot{Q}_{hk}(t)}{T}&=[w_1p_2(t)-u_1p_1(t)]\log{\left[\frac{(u_2+w_1)w_1}{(u_1+w_2)u_1}\right]}+[w_2p_1(t)-u_2p_2(t)]\log{\left[\frac{(u_1+w_2)w_2}{(u_2+w_1)u_2}\right]}\nonumber\\
 	&= \frac{u_1u_2-w_1w_2}{u_1+u_2+w_1+w_2}\log{\left[\frac{u_1u_2}{w_1w_2}\right]}-\lambda(p_1(0)-p_1^{ss})e^{-\sigma t}
 	\nonumber\\
 	&\xrightarrow{\sigma t\gg 1}(j_+-j_-)\log{\left(\frac{j_+}{j_-}\right)}\geq 0
 	\end{align} 
 	where $\lambda=\left\{(u_1+w_1)\log{\left[\frac{(u_2+w_1)w_1}{(u_1+w_2)u_1}\right]}-(u_2+w_2)\log{\left[\frac{(u_1+w_2)w_2}{(u_2+w_1)u_2}\right]}\right\}$.
 	\begin{align}
 	\dot{e}_p(t)&=[w_1p_2(t)-u_1p_1(t)]\log{\left[\frac{p_2(t)w_1}{p_1(t)u_1}\right]}+[w_2p_1(t)-u_2p_2(t)]\log{\left[\frac{p_1(t)w_2}{p_2(t)u_2}\right]}\nonumber\\
 	&\xrightarrow{\sigma t\gg 1} \frac{u_1u_2-w_1w_2}{u_1+u_2+w_1+w_2}\log{\left[\frac{u_1u_2}{w_1w_2}\right]}=(j_+-j_-)\log{\left(\frac{j_+}{j_-}\right)}\geq 0
 	\end{align}  
 	\begin{align}
 	\frac{\dot{h}_d(t)}{T}&=[w_1p_2(t)-u_1p_1(t)]\log{\left[\frac{w_1}{u_1}\right]}+[w_2p_1(t)-u_2p_2(t)]\log{\left[\frac{w_2}{u_2}\right]}\nonumber\\
 	&\xrightarrow{\sigma t\gg 1} \frac{u_1u_2-w_1w_2}{u_1+u_2+w_1+w_2}\log{\left[\frac{u_1u_2}{w_1w_2}\right]}=(j_+-j_-)\log{\left(\frac{j_+}{j_-}\right)}\geq 0
 	\end{align}  
 	\begin{align}
 	\frac{\dot{f}_d(t)}{T}&=[w_1p_2(t)-u_1p_1(t)]\log{\left[\frac{p_2(t)p_1^{ss}}{p_1(t)p_2^{ss}}\right]}+[w_2p_1(t)-u_2p_2(t)]\log{\left[\frac{p_1(t)p_2^{ss}}{p_2(t)p_1^{ss}}\right]}\nonumber\\
 	&=\sigma(p_2(t)p_1^{ss}-p_1(t)p_2^{ss})\log{\left[\frac{p_2(t)p_1^{ss}}{p_1(t)p_2^{ss}}\right]}\geq 0\nonumber\\
 	&\xrightarrow{\sigma t\gg 1} 0
 	\end{align}  
 \end{widetext}
 The relaxation time to a steady state (NESS) from an arbitrary state in a far-from-equilibrium is $\sim \sigma^{-1}=(u_1+u_2+w_1+w_2)^{-1}$, and it is noteworthy that 
 the entropy production inside the system ($T\dot{e}_p$), 
 the total heat production that will be discharged to the surrounding ($\dot{h}_d$), and the housekeeping heat ($\dot{Q}_{hk}$) are all identical at the steady state as $T\dot{e}_p=\dot{h}_d=\dot{Q}_{hk}\rightarrow (j_+-j_-)\log{(j_+/j_-)}\geq 0$, and $\dot{f}_d=0$. 
 Here, $\dot{Q}_{ex}$, the residual of heat (excess heat, $\dot{Q}_{ex}=\dot{h}_d-\dot{Q}_{hk}$) for the nonequilibrium process, is zero at the steady state. 
 Although obtained for 2-state model, the above expression, especially the total heat production (or housekeeping heat) at the steady state, $\dot{h}_d=T\dot{e}_p=\dot{Q}_{hk}=T(j_+-j_-)\log{j_+/j_-}$ can easily be generalized for the $N$-state model. 

\section{5. Relationship between motor diffusivity and heat dissipation.}
 For $(N=2)$-state model one can obtain an explicit expression that relates $D$ with $\dot{Q}$ (for the case of $f=0$) as follows. From the expressions of $V$ (Eq.\ref{eqn:V2}), $D$ (Eq.\ref{eqn:D2}), and $\dot{Q}$, 
 \begin{equation}
 \dot{Q}=\frac{V}{d_0}k_BT\log{\frac{u_1u_2}{w_1w_2}}
 \label{eqn:Q2}
 \end{equation}
 Substitution of $u_1=u_1(V)$ from Eq.\ref{eqn:V2} into Eq.\ref{eqn:Q2} gives an expression of $\dot{Q}$ as a function of $V$: 
 \begin{align}
 \frac{\dot{Q}}{k_{cat}k_BT}&=v\log{\left[\frac{1+\kappa v}{1-v}\right]}\nonumber\\
 &=\sum_{n=1}^{\infty}\frac{1}{n}\left(1+(-1)^{n-1}\kappa^n\right)v^{n+1}
 \label{eqn:Q_v}
 \end{align}
 where $v=V/V_{max}$ ($0\leq v\leq 1$) and $\kappa\equiv\frac{k_{cat}(k_{cat}+w_1+w_2)}{w_1w_2}$. 
 $\dot{Q}$ diverge as $v\rightarrow 1$; but for small $v\ll1$, $\dot{Q}/(k_{cat} k_B T) \sim (\kappa+1)v^2$, thus $v\sim \dot{Q}^{1/2}$. 
 
 As long as $\dot{Q}$ is small, one should expect from Eq.\ref{eq:D_V_2cycle} that $D$ increases with $\dot{Q}$ as  
 \begin{align}
 D=D_0+\gamma_1\dot{q}^{1/2}+\gamma_2\dot{q}+\gamma_3\dot{q}^{3/2}
 \label{eqn:D_q}
 \end{align}
 where $\dot{q}\equiv \frac{\dot{Q}}{k_BTk_{cat}}$, 
 $D_0=\frac{d_0^2k_{cat}}{\kappa+1}$, 
 $\gamma_1=d_0^2k_{cat}\frac{\kappa-1}{2 (\kappa+1)^{3/2}}$, 
 $\gamma_2=-d_0^2k_{cat}\frac{k_{cat}^2}{w_1w_2}$, and 
 $\gamma_3=\frac{d_0^2k_{cat}}{(\kappa+1)^{5/2}}\frac{k_{cat}^2}{w_1w_2}$.

 	For arbitrary number of states $N$, by using Eq.\ref{eq: V_Derrida} and Eq.\ref{eq:V(u_1)}, $\dot{Q}$ can be written as 
 		\begin{align}
 		\dot{Q}/k_BT & = \frac{V}{d_0} \log{\frac{ \prod_{i=1}^N u_i }{ \prod_{i=1}^N w_i } } \nonumber\\
 		& = \frac{V^{(D)}}{N} \log{
 			\left( 1 + V^{(D)} \frac{ \sum_{n=1}^N r_n }{ N }\frac{ \prod_{i=1}^N u_i }{ \prod_{i=1}^N w_i }
 			\right)
 		} \nonumber\\
 		& = \frac{V^{(D)}}{N} \log{
 			\left( 1 + \frac{V^{(D)}}{N} \left( A + B u_1 \right) \frac{ 1 }{C }
 			\right)
 		} \nonumber\\
 		&=\frac{V^{ (D) }}{N} \log{ \left( 1 + \frac{ V^{(D)} }{N} f(V^{ (D) }/N ) \right) }
 		\end{align}
 	where we used $V^{ (D) } = V N / d_0 $, $f(V^{ (D) }/N ) =\frac{A}{C} + B  \left( 1 - B \frac{V^{ (D) } }{N} \right)^{-1}  \left( 1 + \frac{A}{C} \frac{ V^{ (D) }}{N} \right)
 	$, and Eqs.\ref{eq: V_Derrida}, \ref{eq:V(u_1)}, \ref{eq:u_1(V)}.   
 	The definitions of $A, B$, and $C$ are identical to those in Eq.\ref{eq:V(u_1)}.
 	Here $V^{ (D) } / N$ corresponds to ATP hydrolysis rate.
 	For $V^{ (D) } \rightarrow 0$, $\dot{Q} \rightarrow 0$ is expected.
 	Also for small $V^{ (D) }$, $\dot{Q} \sim ( \frac{A}{C} + B ) \left(\frac{V^{(D)}}{N} \right)^2$.
 	Thus, $V \sim \dot{Q}^{1/2}$.  
 	Since $D\sim V+\mathcal{O}(V^2)$ for small $V$, it follows that $D$ can be written as a function of $\dot{q}$ as in the same form as Eq.\ref{eqn:D_q}.

\section{6. Rate constants, enhancement of diffusion, and conversion efficiency determined from the (N=2)-state kinetic model.}
The values in the Table 1 were compiled based on the followings.  
	
	\subsection{Catalase} In Ref.\cite{Riedel2015_Nature} $(\Delta D / D_0)=0.28$ at $ V = 1.7 \times 10^4$ $s^{-1}$; however, $V= 1.7 \times 10^4$ $s^{-1}$ is not the maximum catalytic rate.  
	Because $\Delta D/D_0$ is approximately linear in $V$, the enhancement of diffusion at the maximal turnover rate $V_{\text{max}}=u_2= 5.8 \times 10^4$ is estimated as $(\Delta D / D_0)_{\text{obs}} = 0.28 \times \frac{5.8 \times 10^4}{1.7 \times 10^4} = 0.96 \approx 1$.
	\subsection{Alkaline phosphatase}  
	Similar to catalase, 
	$(\Delta D / D_0)_{\text{obs}} = 0.77 \times \frac{1.4 \times 10^4}{5.5 \times 10^3} = 2.5 \approx 3$.
	\\
	
	\subsection{Estimate of $(\Delta D/D_0)_{\text{max}}$}
	Freely diffusing enzymes effectively perform no work on the surrounding environment; thus $-\Delta \mu_{\text{eff}}=Q$ with $W=0$, which leads to $e^{Q/k_B T}= u_1 u_2/w_1 w_2$.   
	By assuming that the substrate concentration $[S] \sim K_M=(u_2 +w_1) / u_1^o$, we get
    \begin{equation*}
    \begin{aligned}
	 	e^{Q/k_B T}&= \frac{ u_1 u_2 }{w_1 w_2}\
					 	  \sim \frac{ u_1^o K_M u_2 }{w_1 w_2} 
					 	  = \frac{ (u_2 + w_1) u_2 }{w_1 w_2}  \\
					 	  &\geq \frac{  u_2^2 }{w_1 w_2} 
					 	  = \frac{  k_{cat}^2 }{w_1 w_2}. 
	\end{aligned}
	\end{equation*}
	This relation allows us to estimate the upper bound of $(\Delta D/D_0)_{\text{max}}$ as follows when $u_2 \gg w_1, w_2$ is satisfied.   
	\begin{align}
	\left(\frac{\Delta D}{D_0}\right)_{\text{max}}\approx \frac{  k_{cat}^2 }{2 w_1 w_2} \leq \frac{1}{2}e^{Q/k_BT}. 
	\end{align}
	
	Alternatively, $u_1$ and $w_2$ of enzymes can be estimated by assuming (i) that the reaction is diffusion limited, $u_1^o = 10^8 s^{-1} M^{-1}$, and (ii) that the substrate concentration $[S]$ is similar to Michaelis-Menten constant $K_M$ ($[S]\sim K_M$).   
	The two conditions $u_1 = u_1^o [S] \sim K_M \times 10^8$ ($s^{-1}$) and $K_M(=(u_2+w_1)/u_1^o)$, 
	and $Q$ (heat measured by the calorimeter in ref. \cite{Riedel2015_Nature}), $u_2$, $K_M$ which are available in ref. \cite{Riedel2015_Nature}, provide all the rate constants including $w_1=u_1^oK_M-u_2$ and $w_2=\frac{u_1u_2}{w_1e^{Q/k_BT}}$, allowing us to calculate $\left(\frac{\Delta D}{D_0}\right)_{\text{max}}=\frac{u_2^2+(w_1+w_2)u_2-w_1w_2}{2w_1w_2}$. 

\begin{figure*}[ht]
	\centering 
	\includegraphics[scale=0.45]{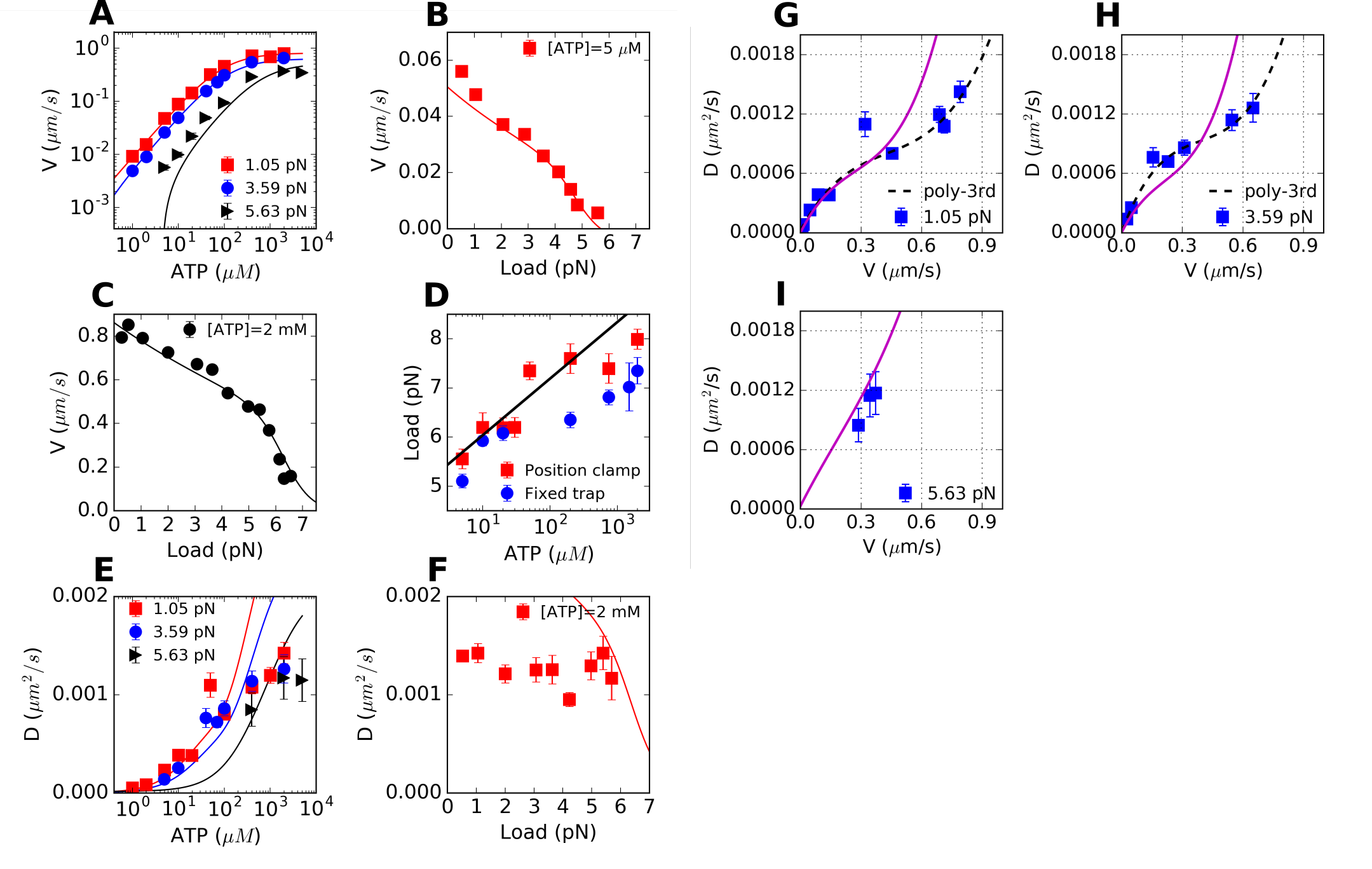}
	
	\caption{
		Analysis of experimental data, extracted from Ref. \cite{Visscher1999_Nature}, but using ($N$=2)-state model.
		The solid lines are the fits to the data
		{\bf A.} $V$ vs ATP at $f=$1.05 pN (red square), 3.59 pN (blue circle), and 5.63 pN (black triangle).
		{\bf B.} $V$ vs load at [ATP] = 5 $\mu$M.
		{\bf C.} $V$ vs load at [ATP] = 2 mM.
		{\bf D.} Stall force as a function of [ATP], measured by `Position clamp' (red square) or `Fixed trap' (blue circle) methods.
		{\bf E.} $D$ vs ATP at $f=$1.05 pN (red square), 3.59 pN (blue circle), and 5.63 pN (black triangle).
		$D$ was estimated from $r=2D/Vd_0$. 
		{\bf F.} $D$ vs load at [ATP] = 2 mM.
		{\bf G-I.} Motor diffusivity ($D$) as a function of mean velocity ($V$) for kinesin-1.  
		($V$,$D$) measured at varying [ATP] (= 0 -- 2 mM) and a fixed ({\bf G}) $f=$1.05 pN, ({\bf H}) 3.59 pN, and ({\bf I}) 5.63 pN \cite{Visscher1999_Nature}. 
		The black dashed lines in {\bf G} and {\bf H} are the fits using Eq.\ref{eq:D_V}.
		The solid lines in magenta in {\bf G-I} are plotted using the ($N$=2)-kinetic model.
	}
	\label{fig_kinesin_refit_fix_2states}
\end{figure*}

\end{document}